\documentclass[12pt]{article}
\pdfoutput=1

\usepackage[english]{babel}
\usepackage[latin1]{inputenc}

\usepackage[intlimits]{amsmath}
\usepackage{amssymb}

\usepackage{url}
\usepackage{graphicx}

\def\bea{\begin{eqnarray}}
\def\eea{\end{eqnarray}}

\def\1{{\bf 1}}

\def\bo{{\raise.15ex\hbox{\large$\Box$}}}               % D'Alembertian
\def\face{{\raise.2ex\hbox{$\displaystyle \bigodot$}\mskip-2.2mu \llap {$\ddot
        \smile$}}}                                      % happy face
\def\leftrightarrowfill{$\mathsurround=0pt \mathord\leftarrow \mkern-6mu
        \cleaders\hbox{$\mkern-2mu \mathord- \mkern-2mu$}\hfill
        \mkern-6mu \mathord\rightarrow$}       % <--> double differential
\def\dvec#1{\vbox{\ialign{##\crcr
        \leftrightarrowfill\crcr\noalign{\kern-1pt\nointerlineskip}
        $\hfil\displaystyle{#1}\hfil$\crcr}}}           % <--> accent
\def\beq{\begin{equation}}
\def\eeq{\end{equation}}

\def\lsim{\mathrel{\raise.3ex\hbox{$<$\kern-.75em\lower1ex\hbox{$\sim$}}}}
\def\gsim{\mathrel{\raise.3ex\hbox{$>$\kern-.75em\lower1ex\hbox{$\sim$}}}}

\catcode`@=12
\begin{document}
\date{\mbox{ }}

\title{ 
{\normalsize     
DESY 09-062\hfill\mbox{}\\
August 2009\hfill\mbox{}\\}
\vspace{2cm}
\bf Gravitino Dark Matter\\ and general neutralino NLSP  \\[8mm]}
%
%\vspace{2cm} 
\author{L.~Covi$^a$, J. Hasenkamp$^a$, S. Pokorski$^b$ and
J. Roberts$^c$\\[2mm]
{\small\it a Deutsches Elektronen-Synchrotron DESY, Hamburg, Germany}\\
{\small\it b Institute of Theoretical Physics, Warsaw university, Warsaw, Poland}\\
{\small\it c Center for Cosmology and Particle Physics, 
New York University, New York, NY 10003, USA}
}
\maketitle

\thispagestyle{empty}

\vspace{1cm}
\begin{abstract}
\noindent
We study the scenario of gravitino DM with a general neutralino NLSP
in a model independent way. We consider all neutralino decay channels
and compare them with the most recent BBN constraints.  We check how
those bounds are relaxed for a Higgsino or a Wino NLSP in comparison
to the Bino neutralino case and look for possible loopholes in the
general MSSM parameter space.  We determine constraints on the
gravitino and neutralino NLSP mass and comment on the possibility of
detecting these scenarios at colliders.
\end{abstract}

\newpage

\section{Introduction}

The gravitino was the first supersymmetric Dark Matter (DM)
candidate \cite{gravitinoDM} and is still one of the best motivated
today. In some of the supersymmetry breaking schemes, such as 
those that rely on gauge~\cite{GMSB} or even gaugino 
mediation~\cite{gauginoMSB},  such a state is either automatically,
or in the second case very naturally~\cite{gravitinogauginomed},
the lightest supersymmetric particle.  
An equally important motivation is thermal leptogenesis: 
producing the lepton
asymmetry without stretching parameters requires a relatively high
reheating temperature around $10^{10}$~GeV~\cite{leptogenesis}, which
cannot be reconciled with an unstable gravitino unless it is very
heavy or so light as to be irrelevant for the energy budget of the
universe.  In general the gravitino energy density from thermal
scatterings depends on the supersymmetric parameters and the
temperature as~\cite{GravitinoY}
\begin{equation}
\Omega_{3/2} h^2 \simeq 0.3  
\left(\frac{100\;\mbox{GeV}}{ m_{3/2}}\right)
\left(\frac{m_{\tilde g} }{ 1\;\mbox{TeV} }\right)^2 
\left(\frac{T_R }{ 10^{10}\;\mbox{GeV}}\right)\; .
\end{equation}
This result arises from the fact that the gravitino interactions in
the goldstino limit are inversely proportional to its mass and
therefore larger gravitino masses correspond to more weakly
interacting gravitinos and thus lower number densities.  Therefore a
heavy stable gravitino does allow for a large reheating temperature,
but it also implies a long lifetime for the
Next-to-Lightest-Supersymmetric-Particle (NLSP) if R-parity is
conserved. This in turn risks a clash between NLSP decay and Big Bang
Nucleosynthesis (BBN)~\cite{BBN-gravitino}. 
In this work we will investigate which mass ranges are still consistent 
with a gravitino LSP as DM for a general neutralino NLSP.

Specific models like the CMSSM have been studied in \cite{CMSSM} and
the case of a general Bino NLSP has been considered with some
approximations in~\cite{Feng:2004mt, Kawasaki:2008qe}; in these
papers it was found that it is very difficult to reconcile such
scenarios with a gravitino mass above $1 $ GeV or with the
parametric equality $ m_{3/2} = m_0 $.  We would like to see if this
is true in the more general case. For a non-Bino neutralino we would
naively expect the situation to improve due to the smaller number
densities or due to the different branching ratios into hadrons.

The case of a Higgsino NLSP naturally has a larger annihilation
cross-section, leading to smaller number densities. However in the
CMSSM it arises only in the finely tuned focus-point region, which
corresponds to very large supersymmetric masses. If we relax some of
the usual assumptions on universality, we can obtain a Higgsino NLSP
at lower masses, in particular in more general models of gauge
mediation of supersymmetry breaking~\cite{higgsino-gaugemed}.  A
partially Higgsino NLSP can also be found in models with non-universal 
Higgs masses, such as in gaugino mediation \cite{gauginomed}.

A Wino (N)LSP has predominantly been studied in the context of anomaly
mediation~\cite{anomalymed}, where it is naturally one of the lightest
particles. Unfortunately in those models the gravitino is always very 
heavy and cannot be the LSP. We will consider here the possibility 
of a Wino NLSP for the most general SUSY breaking parameters and 
in particular for low values of the gaugino mass parameter $ M_2 $ 
and non universal boundary conditions. A Wino NLSP can occur naturally 
in certain GUT models with F-terms which are non-singlet under
the GUT group~\cite{WinoNLSP}. Note that a Wino NLSP may be 
possible also in the case of $U(1)'$ mediated supersymmetry 
breaking~\cite{Langacker:2008ip}. 

The goal of this analysis is to close any gap in the study of the
neutralino NLSP with gravitino DM scenario, either coming from
simplifications in the computations of the NLSP number density
or from approximations in its decay rates and hadronic branching
ratio. For this purpose we compute and give the complete results 
for the two and three body neutralino decays.

This paper is organised as follows: after summarising the general
neutralino properties in Section~2, we will discuss its decays 
into gravitino and Standard Model particles in Section~3, assuming 
it is the NLSP. We will discuss with particular care the hadronic
channels since they tend to be more strongly constrained by
Nucleosynthesis. We will then review the computation of the 
neutralino relic abundance in Section 4 and then consider specifically 
the cases of Bino-Wino, Bino-Higgsino and Wino-Higgsino neutralino 
and compare them directly with the BBN constraints of
reference~\cite{Jedamzik}.
Furthermore we will generalise our result for any neutralino mixing 
in Section 5 and draw our conclusions concerning supersymmetry 
breaking parameters and specific models in Section 6.

\medskip 

\section{Neutralino mass matrix and SUSY breaking\\ parameters}

Supersymmetry and EW symmetry breaking produce a non-trivial mass
matrix for the neutralino in the basis of the gaugino-Higgsino states,
given in the Bino-Wino-Higgsino basis by \cite{SUSYprimer}
\begin{equation}
M = \left(
\begin{array}{cccc}
M_1 & 0 & - M_Z s_W c_\beta & M_Z s_W s_\beta \cr
0  & M_2 &  M_Z c_W c_\beta &  - M_Z c_W s_\beta \cr
- M_Z s_W c_\beta & M_Z c_W c_\beta & 0 & \mu \cr
M_Z s_W s_\beta & - M_Z c_W s_\beta & \mu & 0 
\end{array} \right)\; ,
\end{equation}
where $M_1, M_2 $ are the supersymmetric gaugino mass parameters
for the $U(1)$ and $SU(2)$ sector respectively, $\mu $ is the
supersymmetric Higgs mass parameter, $M_Z$ the mass of the
Z gauge boson, $ s/c_W = \sin/\cos\theta_W $ with $\theta_W$ 
being the Weinberg angle, while  $ s/c_\beta = \sin/\cos\beta $ 
is determined by the ratio of the Higgs {\it v.e.v.}s as
$\tan\beta = v_u/v_d $.

The mass eigenstates are mixed gaugino/Higgsino states and their
mixing angles and masses are determined by the SUSY breaking
parameters $ M_1, M_2$ and $\tan\beta, \mu $. In the following we will 
consider these as free parameters. In general then the mass matrix 
above can be diagonalised by a unitary matrix $N_{ij} $ such that
\begin{equation}
\chi^0_i = N_{ij} \Psi^0_j
\end{equation}
where $ \Psi^0 = (\tilde B, \tilde W^0, \tilde H_d^0, \tilde H_u^0 )^T
$.  If we consider all the parameters in the mass matrix to be real,
as we will do here, then the matrix $ N $ can be chosen real and
orthogonal.  The particle we are interested in is the lightest
neutralino $\chi_1^0 $, which may naturally become the NLSP in many
supersymmetry breaking scenarios with a gravitino LSP.  As is clear
from the mass matrix above, the values of the parameters $ M_1, M_2,
\mu $ (and more weakly $\tan\beta $) determine the composition of the
lightest neutralino and therefore its decay channels and relic
abundance.  In the following we will take the neutralino mass
parameters at the low scale as the inputs and not assume any special
relation between them.  In this way we will be able to explore the
most general scenario with a neutralino NLSP. But first we discuss
the decay channels for the neutralino gauge eigenstates.

\medskip

\section{Neutralino NLSP decays}

In the case of conserved R-parity and {\bf a} gravitino LSP, the
neutralino couples with gravitino and gauge boson or Higgs boson via
the supergravity interactions~\cite{JH}.  In general for gravitino LSP
the dominant couplings are those involving the Goldstino component of
the gravitino and therefore they are enhanced by 
$m_{\tilde \chi}/m_{\frac{3}{2}}$.  
Therefore the neutralino lifetime always
contains an overall factor $ x_{\frac{3}{2}}^2 = (m_{\frac{3}{2}}/
m_{\tilde \chi})^2 $.  In the following we will discuss first the
decays for the neutralino pure states: Bino, Wino and Higgsino.

\subsection{Primary Bino decay channels}

A Bino neutralino can decay into a photon and a gravitino via 
its photino component and the decay rate is given by
\begin{eqnarray}
\Gamma \left(\tilde B \rightarrow \psi_{3/2}\; \gamma \right)
&=& 
\frac{\left| \cos \theta_W  \right|^2}{48 \pi M_P^2} 
\frac{m_{\tilde B}^3}{x_{\frac{3}{2}}^2} 
\left(  1-x_{\frac{3}{2}}^2 \right)^3 \left( 1+3 x_{\frac{3}{2}}^2 \right) ,
 \end{eqnarray}
 where $ \theta_W $ is the Weinberg angle, $ m_{\tilde B} $ the
 Bino mass. For a negligible gravitino mass in the phase space factor, 
the decay of a pure Bino state is dominated by this channel and 
the lifetime can be approximated as:
\begin{eqnarray}
\Gamma^{-1} \left(\tilde B \rightarrow \psi_{3/2}\; \gamma \right)
&=& 7.7 \times 10^4 \mbox{s} 
\left(\frac{m_{\tilde B}}{100\; \mbox{GeV}}\right)^{-5}
\left(\frac{m_{\frac{3}{2}}}{1\; \mbox{GeV}} \right)^2\; .
 \end{eqnarray}

\begin{figure}%[t]
\centering
\includegraphics[scale=0.8]{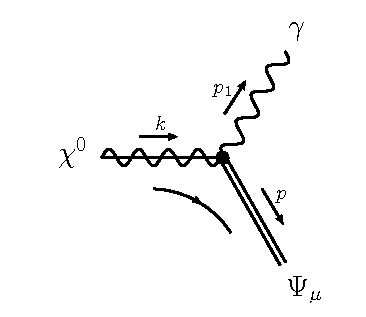}
\includegraphics[scale=0.8]{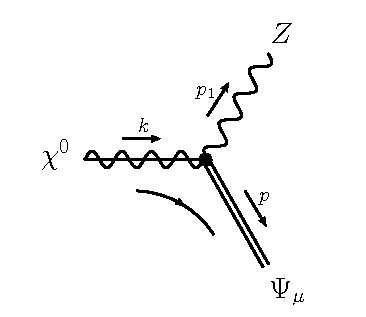}
\caption{\small 
Feynman diagrams for the decay of a gaugino neutralino
into neutral gauge bosons and gravitino.
}
\label{neutralino-gauge}
\end{figure}

If it is kinematically allowed, the Bino can also decay into 
a Z boson and a gravitino, the decay rate for this channel is given by
\begin{eqnarray}
\Gamma \left( \tilde B \rightarrow \psi_{3/2} Z \right) 
&=& \frac{\left| \sin \theta_W \right|^2}{48 \pi M_P^2} 
\frac{m_{\tilde B}^3}{x_{\frac{3}{2}}^2} 
\sqrt{1- 2(x_{\frac{3}{2}}^2 +x_{Z}^2) + (x_{Z}^2 - x_{\frac{3}{2}}^2)^2} 
\nonumber \\ 
&& \times \lbrack \left( 1-x_{\frac{3}{2}}^2 \right)^2 \left( 1+3 x_{\frac{3}{2}}^2 \right) -x_{Z}^2 \lbrace 3+x_{\frac{3}{2}}^3 \left( x_{\frac{3}{2}} -12 \right) \nonumber \\ 
&& - x_{Z}^2 \left( 3 
- x_{\frac{3}{2}}^2 - x_{Z}^2 \right)  \rbrace \rbrack 
\end{eqnarray}
where here $ x_Z = M_{Z}/m_{\tilde B} $.

So for large Bino masses the Bino lifetime is approximately given by
\begin{eqnarray}
\Gamma^{-1} \left(\tilde B \rightarrow \psi_{3/2}\; \gamma/Z \right)
&=& 57\; \mbox{s} 
\left(\frac{m_{\tilde B}}{1\;\mbox{TeV}}\right)^{-5}
\left(\frac{m_{\frac{3}{2}}}{10\;\mbox{GeV}} \right)^2\; ,
\label{Bino-lifetime}
 \end{eqnarray}
so that it becomes shorter than 0.1 s for masses larger than
$ 3.6\; \mbox{TeV} 
\times \left(\frac{m_{\frac{3}{2}}}{10\;\text{\small GeV}} \right)^{2/5} $.

\subsection{Primary Wino decay channels}

A Wino neutralino can also decay into a photon and a gravitino via 
its photino component and the decay rate is given in this case by
\begin{eqnarray}
\Gamma \left(\tilde W \rightarrow \psi_{3/2}\; \gamma \right)
&=& 
\frac{\left| \sin \theta_W  \right|^2}{48 \pi M_P^2} 
\frac{m_{\tilde W}^3}{x_{\frac{3}{2}}^2} 
\left(  1-x_{\frac{3}{2}}^2 \right)^3 \left( 1+3 x_{\frac{3}{2}}^2 \right) ,
 \end{eqnarray}
 where $ m_{\tilde W} $ is the Wino mass.

As with the Bino, the lifetime of a pure Wino state is dominated by
this channel and for a negligible gravitino mass in the phase space
factor it can be simplified to:
\begin{eqnarray}
\Gamma^{-1} \left(\tilde W \rightarrow \psi_{3/2}\; \gamma \right)
&=& 2.5\;\times 10^5\; \mbox{s} 
\left(\frac{m_{\tilde W}}{100\;\mbox{GeV}}\right)^{-5}
\left(\frac{m_{\frac{3}{2}}}{1\;\mbox{GeV}} \right)^2\; .
 \end{eqnarray}
 
For a heavier Wino the channel into a Z boson and a gravitino 
opens up, with the decay rate
\begin{eqnarray}
\Gamma \left( \tilde W \rightarrow \psi_{3/2} Z \right) 
&=& \frac{\left| \cos \theta_W \right|^2}{48 \pi M_P^2} 
\frac{m_{\tilde W}^3}{x_{\frac{3}{2}}^2} 
\sqrt{1- 2(x_{\frac{3}{2}}^2 +x_{Z}^2) + (x_{Z}^2 - x_{\frac{3}{2}}^2)^2} 
\nonumber \\ 
&& \times \lbrack \left( 1-x_{\frac{3}{2}}^2 \right)^2 \left( 1+3 x_{\frac{3}{2}}^2 \right) -x_{Z}^2 \lbrace 3+x_{\frac{3}{2}}^3 \left( x_{\frac{3}{2}} -12 \right) \nonumber \\ 
&& - x_{Z}^2 \left( 3 
- x_{\frac{3}{2}}^2 - x_{Z}^2 \right)  \rbrace \rbrack ,
\end{eqnarray}
similar to the case of a Bino neutralino.
We see clearly that the decay into Z is stronger for the Wino case 
than for the Bino and this immediately gives a stronger decay into 
hadrons. For a heavy Wino, the decay rate becomes equal to the Bino one,
 as given in eq.~(\ref{Bino-lifetime}).

For large Wino masses the channel into a gravitino and
a W pair opens up, via the Wino-gravitino-W-W 4-vertex, which 
arises from the non-abelian interaction, in addition to the corresponding 
three gauge bosons vertex and other diagrams with intermediate
charginos and Higgs bosons. The main part of the contribution
from the gauge sector, which appears at first sight to be enhanced 
by power of $ m_{\tilde W}^4/M_W^4 $ compared to the 2-body decays, 
cancels out thanks to gauge invariance, as found also in 
\cite{Ferrantelli:2007bx}. We also found a cancellation of the
sub-leading terms $ m_{\tilde W}^2/M_W^2 $, so this channel, suppressed 
by an additional $\alpha_{em}$ and the 3-body phase space always remains 
negligible.

\subsection{Primary Higgsino decay channels}

A pure Higgsino has direct decay channels into the scalar and
pseudo-scalar Higgses and a gravitino, as well as into a Z boson and
gravitino. In case the bosons are too heavy to be produced on-shell,
 the decay proceeds via off-shell Higgses mainly into heavy quarks, 
 $b \bar{b} $, whilst the decay through an off-shell Z goes to both 
quarks and leptons.
 The decay rate into an on-shell Z boson is given by
\begin{eqnarray}
\Gamma \left( \tilde H \rightarrow \psi_{3/2} Z \right) 
&=& \frac{\left| - N_{13} \cos \beta + N_{14} \sin\beta \right|^2}{96 \pi M_P^2} 
\frac{m_{\tilde H}^3}{x_{\frac{3}{2}}^2} 
\sqrt{1- 2(x_{\frac{3}{2}}^2 +x_{Z}^2) + (x_{Z}^2 - x_{\frac{3}{2}}^2)^2} 
\nonumber \\ 
&& \times \lbrack \left( 1+x_{\frac{3}{2}} \right)^2 
\left( 1- x_{\frac{3}{2}} \right)^4 
-x_{Z}^2 \lbrace  \left( 1-x_{\frac{3}{2}}\right)^2 
\left(3+2 x_{\frac{3}{2}} - 9 x_{\frac{3}{2}}^2 \right)\nonumber \\ 
&& - x_{Z}^2 \left( 3 
- 2 x_{\frac{3}{2}} - 9 x_{\frac{3}{2}}^2 - x_{Z}^2 \right)  \rbrace \rbrack , 
\label{Higgsino-Z}
\end{eqnarray}
where $ N_{ij} $ is the neutralino mixing matrix and we consider
a mixed Higgsino initial state, $ m_{\tilde H} $ is the Higgsino mass
and again $ x_a = M_a/m_{\tilde H}$.
The decay into light Higgs is instead 
\begin{eqnarray}
\Gamma \left( \tilde H \rightarrow \psi_{3/2} h \right) 
&=& \frac{\left| - N_{13} \sin \alpha + N_{14} \cos\alpha \right|^2}{96 \pi M_P^2} 
\frac{m_{\tilde H}^3}{x_{\frac{3}{2}}^2} 
\sqrt{1- 2(x_{\frac{3}{2}}^2 +x_{h}^2) + (x_{h}^2 - x_{\frac{3}{2}}^2)^2} 
\nonumber \\ 
&& \times \lbrack \left( 1-x_{\frac{3}{2}} \right)^2 
\left( 1+ x_{\frac{3}{2}} \right)^4 
-x_{h}^2 \lbrace  \left( 1+x_{\frac{3}{2}}\right)^2 
\left(3- 2 x_{\frac{3}{2}} + 3 x_{\frac{3}{2}}^2 \right)\nonumber \\ 
&& - x_{h}^2 \left( 3 + 2 x_{\frac{3}{2}} + 3 x_{\frac{3}{2}}^2 - x_{h}^2 \right)  \rbrace \rbrack , 
\label{Higgsino-h}
\end{eqnarray}
where $\alpha $ is the mixing angle between the two real parts of the 
Higgs bosons $ H_u $ and $ H_d $ into the $h$ and H mass eigenstates. 

\begin{figure}[t]
  \centering
  \includegraphics[scale=0.8]{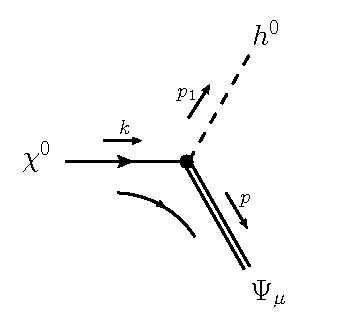}
  \includegraphics[scale=0.8]{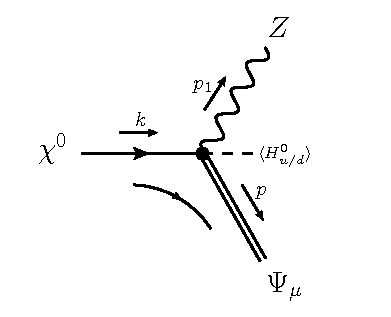}
  \caption{\small 
    Feynman diagrams for the decay of a Higgsino neutralino
    into neutral Higgs or gauge bosons and gravitino.
  }
  \label{higgsino-boson}
\end{figure}

For large neutralino masses the decays into the heavy Higgs bosons
open up. The heavy scalar Higgs channel is given by the above expression
(\ref{Higgsino-h}), but with the overall prefactor given by the orthogonal 
combination $\left| N_{13} \cos \alpha + N_{14} \sin\alpha \right|^2 $
and $ x_h \rightarrow x_H $, while the decay to the pseudo-scalar Higgs 
has the same form as the decay into the Z boson in eq.~(\ref{Higgsino-Z}), 
but with the overall prefactor 
$ \left| N_{13} \sin \beta + N_{14} \cos \beta \right|^2 $ and 
$ x_Z \rightarrow x_A $.
We will not consider these decays further since they remain sub-dominant
apart from the case when the neutralino composition is such to suppress 
the light Higgs channel.

In the decoupling limit $ \sin\alpha = - \cos\beta $ and 
$ \cos\alpha = \sin\beta $, and for $ N_{13} = N_{14} $, 
the decay time of a Higgsino neutralino is given by
\begin{eqnarray}
\Gamma^{-1} \left(\tilde H \rightarrow \psi_{3/2}\; h/Z \right)
&=& 114 \; \mbox{s} 
\left(\frac{m_{\tilde H}}{1\;\mbox{TeV}}\right)^{-5}
\left(\frac{m_{\frac{3}{2}}}{10\;\mbox{GeV}} \right)^2\; ,
\label{Higgsino-lifetime}
 \end{eqnarray}
so that it becomes shorter than 0.1 s for masses larger than
$ 4.1\; \mbox{TeV} \times \left(\frac{m_{\frac{3}{2}}}{10\;\mbox{GeV}} \right)^{2/5} $.

Note that for large Higgsino masses a supergravity
4-vertex may become important. In this case it is the
Higgsino-gravitino-$h$-Z 4-vertex, which arises from the Higgs
scalar derivative coupling, and allows the direct decay into
a gravitino, light Higgs and a Z. An analogous final state is
also obtained by Higgs radiation from an intermediate Z,
 Z radiation from a pseudo-scalar Higgs or via neutralino
 intermediate states.
In this case again an important cancellation of terms of
order $ m_{\tilde H}^2/M_Z^2 $ takes place between the different 
diagrams thanks to gauge invariance and so this channel
is always sub-dominant.

\subsection{Gaugino hadronic branching ratios}

\begin{figure}[ht]
  \centering
  \includegraphics[scale=1]{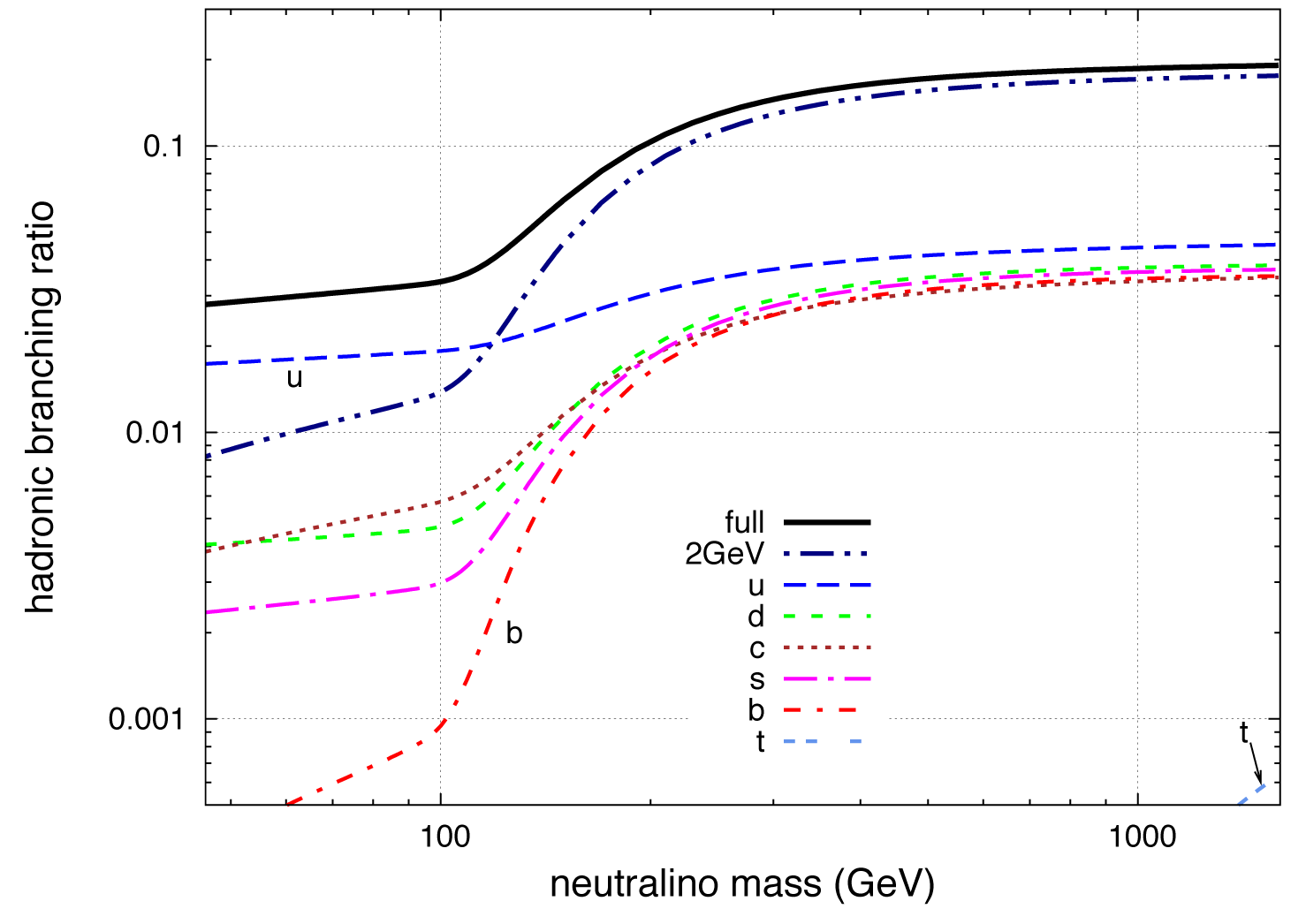}
  \caption{\small 
    Branching ratios into a quark pair and gravitino for a Bino 
    neutralino and negligible gravitino mass in the phase
    space factor. Below the Z threshold the decay is dominated
    by the off-shell photon and there the light quark channels
    are preferred due to the IR logarithmic enhancement.
    Above the Z threshold, the Z channel starts to become important
    and all light quarks are produced democratically.
    The top quark production is always negligible due to the
    strong phase space suppression, even for large neutralino
    masses.
  }
  \label{Bino-quarks}
\end{figure}

The gaugino decay into hadrons proceeds via an intermediate
off-shell photon, via an off-/on-shell Z-boson and also via
intermediate off-shell squarks.
The decay via intermediate photons is particularly simple
in the limit of near massless quarks. For the up 
quark at leading order in $ x_{\frac{3}{2}} $ and $m_u $ this is given by:
\begin{eqnarray}
\Gamma \left( \tilde \chi_G \rightarrow \psi_{3/2} \gamma^*
\rightarrow \psi_{3/2} u \bar u \right) 
&=& 
\frac{\left| N_{11} \cos \theta_W + N_{12} \sin \theta_W  \right|^2}
{27 (2 \pi)^3 M_P^2} 
\frac{m_{\tilde\chi_G}^3}{x_{\frac{3}{2}}^2} 
\log \left(\frac{m_{\tilde\chi_G}}{2 m_u} \right)\; .
\end{eqnarray}
Here we see that the contribution is actually enhanced by a
logarithmic IR divergence. Therefore the branching ratio into quarks
never quite vanishes, always remaining greater than 0.03. This same
logarithmic divergence ensures that the light quarks dominate the
hadronic channels for low gaugino masses. Note that for late times in
Nucleosynthesis, above 100 s or so, it is assumed that only quarks
produced with energy above 2 GeV are able to hadronise into nucleons
and affect the BBN predictions as hadrons, while at lower energies
they can only end up in light mesons that decay before interacting and
produce mostly leptons or photons~\cite{Kohri:2001jx, Kawasaki:2004qu}. 
Therefore in Figure~\ref{Bino-quarks} we also
give the result with a 2 GeV IR cut-off. We see that the introduction
of such a scale only modifies the hadronic branching ratio at low
neutralino masses and reduces it maximally by a factor three compared
to the one computed with the physical quark masses. In general we find
a hadronic branching ratio always much larger than $10^{-3} $, a value
which is often taken as the reference minimal value for a Bino
neutralino.  To be conservative we will use the result obtained with
the physical quark masses in our exclusion plots. Note that in the 
region of the parameter space in which the NLSP has a short lifetime, 
the effect of the interconversion of protons
to neutrons from light mesons can be important~\cite{Kohri:2001jx}.
For the case of an intermediate Z boson the IR divergence is not
present and the contribution becomes largest for an on-shell Z with
the branching ratio being at most equal to the decay into a Z boson
weighted by the respective hadronic width as discussed in
\cite{Feng:2004mt}.

The channel with intermediate squarks has been neglected so far. 
It is clearly negligible for heavy squarks. We have implemented 
the full calculation of this contribution for varying squark masses.
We find that in the limit of large universal squark
masses this contribution reduces to
\begin{eqnarray}
\Gamma \left( \tilde \chi_G \rightarrow \psi_{3/2} \tilde q
\rightarrow \psi_{3/2} q \bar q \right) 
&\rightarrow & 
\frac{\left| N_{11} \cos \theta_W + N_{12} \sin \theta_W  \right|^2}
{27 (2 \pi)^3 M_P^2} 
\frac{m_{\tilde\chi_G}^7}{m_{\tilde q}^4 x_{\frac{3}{2}}^2}   \; ,
\end{eqnarray}
and it is suppressed by at least a factor 
$ \left(m_{\tilde\chi_G}/m_{\tilde q}\right)^4 $ compared to the 
other channels. Surprisingly also for nearly degenerate masses
between the squarks and the neutralino, the contribution of the 
squarks of the third generation amounts only to at most 1\% of 
the hadronic branching ratio, due to the much stronger off-shell 
photon channel in the light quarks~\cite{JH}. 
A larger relative correction of about 7\% arises if the degenerate 
scalar is the superpartner of the up quark and the neutralino mass 
is below the Z threshold, but in that case the hadronic branching
ratio is small, so the increase is anyway negligible. Moreover 
the mass degeneracy has to be pretty extreme, about 1\%, and then
also coannihilation effects between the neutralino and the squarks 
become important. We therefore neglect these terms in the parameter 
scans of Section~\ref{parameterScans}.

We show the result of the full calculation of the branching ratio to
the different quark final states for a pure Bino in
Figure~\ref{Bino-quarks}. We see that, as expected, the light quarks
dominate for small neutralino masses and that the top quark
contribution is always negligible. Note that the hadronic branching
ratio is practically independent of the gravitino mass since there is
an overall factor of $ x_{\frac{3}{2}}^{-2} $ in all the diagrams that
drops out when we calculate the branching ratio. The remaining
dependence on $ x_{\frac{3}{2}} $ from the phase space factor is weak.
The most significant effect of a large gravitino mass is to move the
location of the Z threshold.

\begin{figure}[t]
  \centering
  \includegraphics[scale=1]{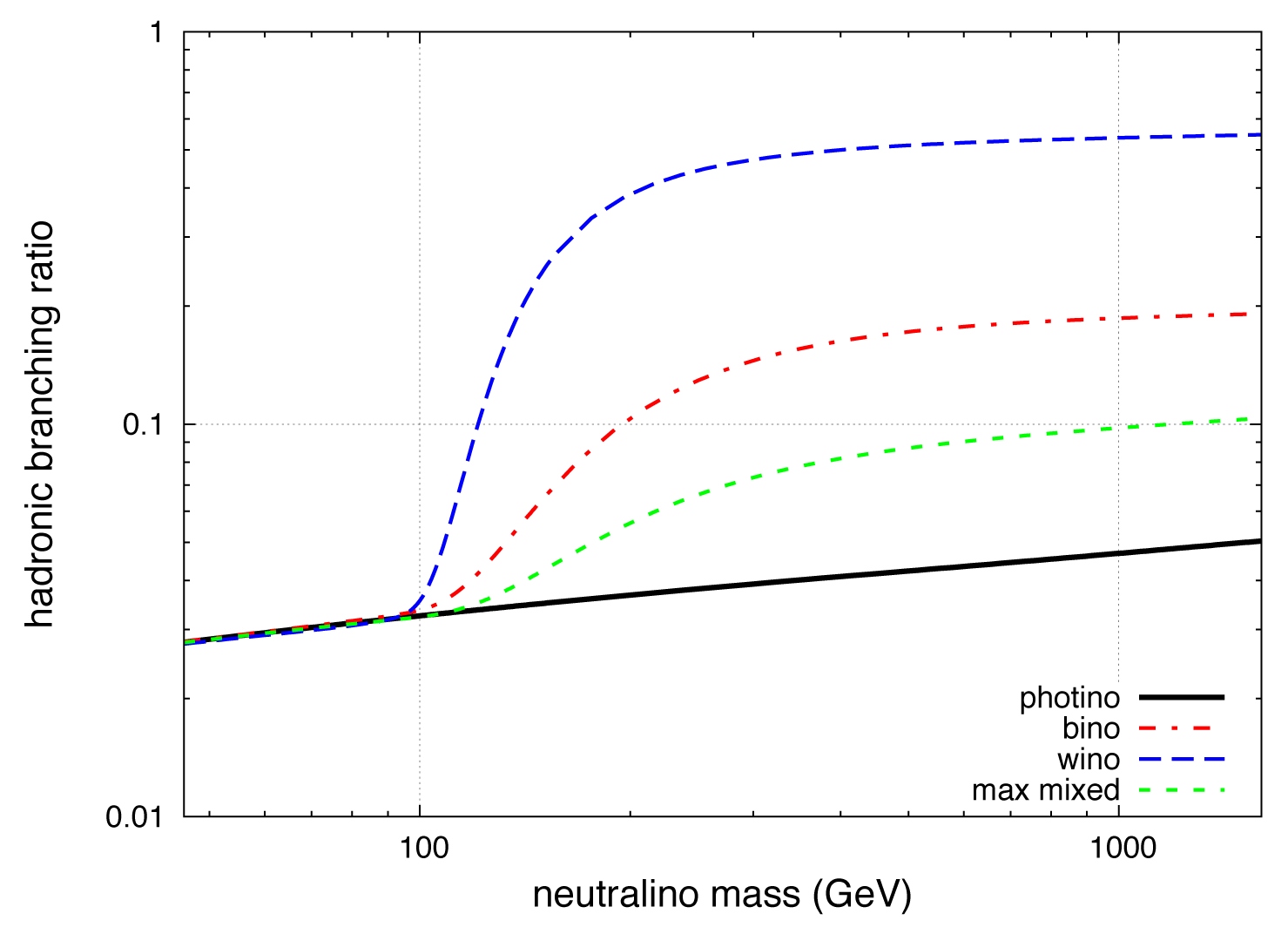}
  \caption{\small Branching ratio into a quark pair and a gravitino for
    a Bino or Wino neutralino, for the maximally mixed state and for a
    photino. Clearly visible is the opening of the Z channel, which 
    increases the hadronic branching ratio for any state not exactly
    equal to the photino mixture.
  }
  \label{BR-gaugino}
\end{figure}

The results for the branching ratios discussed above cannot be
directly applied to the general gaugino or neutralino case, since
interference effects between the different diagrams may play a role
and either increase the decay rate and/or lower the hadronic branching
ratio.  Fortunately the main interference effect arises from the
neutralino composition. One such case happens for a mixed Bino-Wino
state. In this case the branching ratio into quarks is lower than in
either of the pure channels. This happens if the Wino-Bino mixture is
such to as give an approximately photino NLSP. This significantly
reduces the contribution of the Z channel to the total decay rate
whilst boosting the two-body decay into photon and gravitino.  This
effect can keep the hadronic branching ratio small even above the Z
threshold as shown in Figure~\ref{BR-gaugino}.

\subsection{Higgsino hadronic branching ratios}

The Higgsino hadronic decays proceed mainly via the intermediate off-
or on-shell Z and Higgs bosons. Below the Higgs threshold the Z channel 
is dominant and determines the branching ratio to the different quark 
flavours. Above the light Higgs threshold the decay into $ b \bar b $ 
quarks quickly takes over as it is the dominant Higgs decay channel.  
In fact in the MSSM the light Higgs mass is always below the threshold 
for decay into WW and for the heavy Higgses we will consider 
mostly the case of near-decoupling, where they cannot be produced 
on-shell in neutralino decays and their contribution to the
hadronic branching ratio is completely negligible. 

Above the threshold for WW production and below the $t\overline{t}$ 
threshold from Higgs decay, a 5-body hadronic channel opens up for 
the Higgsino. In this case we can consider the (heavy) Higgs as on-shell 
and estimate the contribution from the 5 body-decay using the expression
\begin{eqnarray}
B_{had} &\sim  & B_{had}^{3-body} +
\frac{ \Gamma \left( \tilde H \rightarrow \psi_{3/2} H
\right)}{\Gamma_{tot}} 
\left( 1- (B_{lep}^W)^2 \right) \; .
\end{eqnarray}
where we have taken the branching ratio of a heavy Higgs into a WW pair 
to be one.  We have checked that this additional channel does not
substantially change the higgsino hadronic branching ratio since 
it is already quite large at around 80\%. 

\begin{figure}[t]
  \centering
  \includegraphics[scale=0.61]{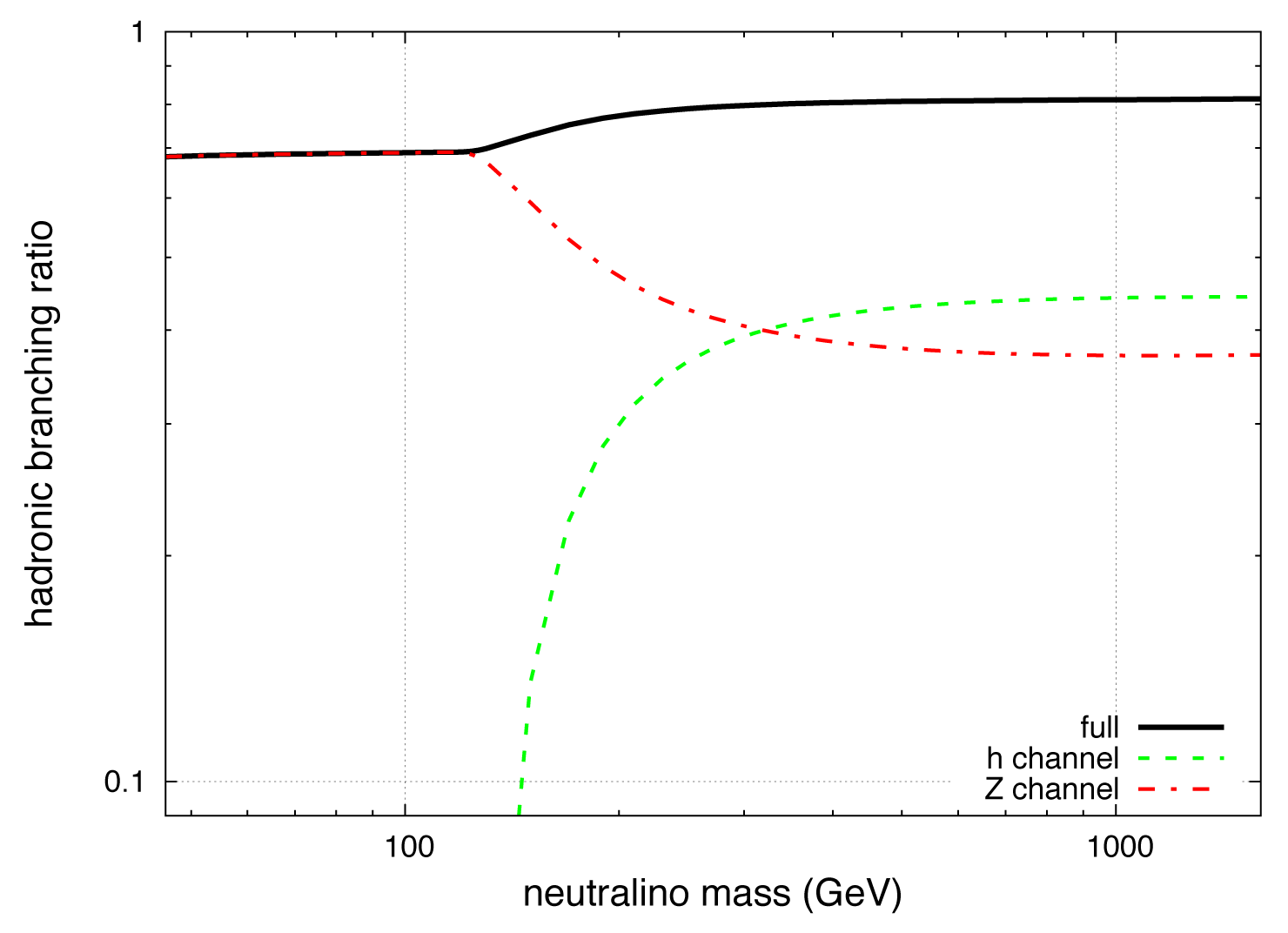}
  \includegraphics[scale=0.61]{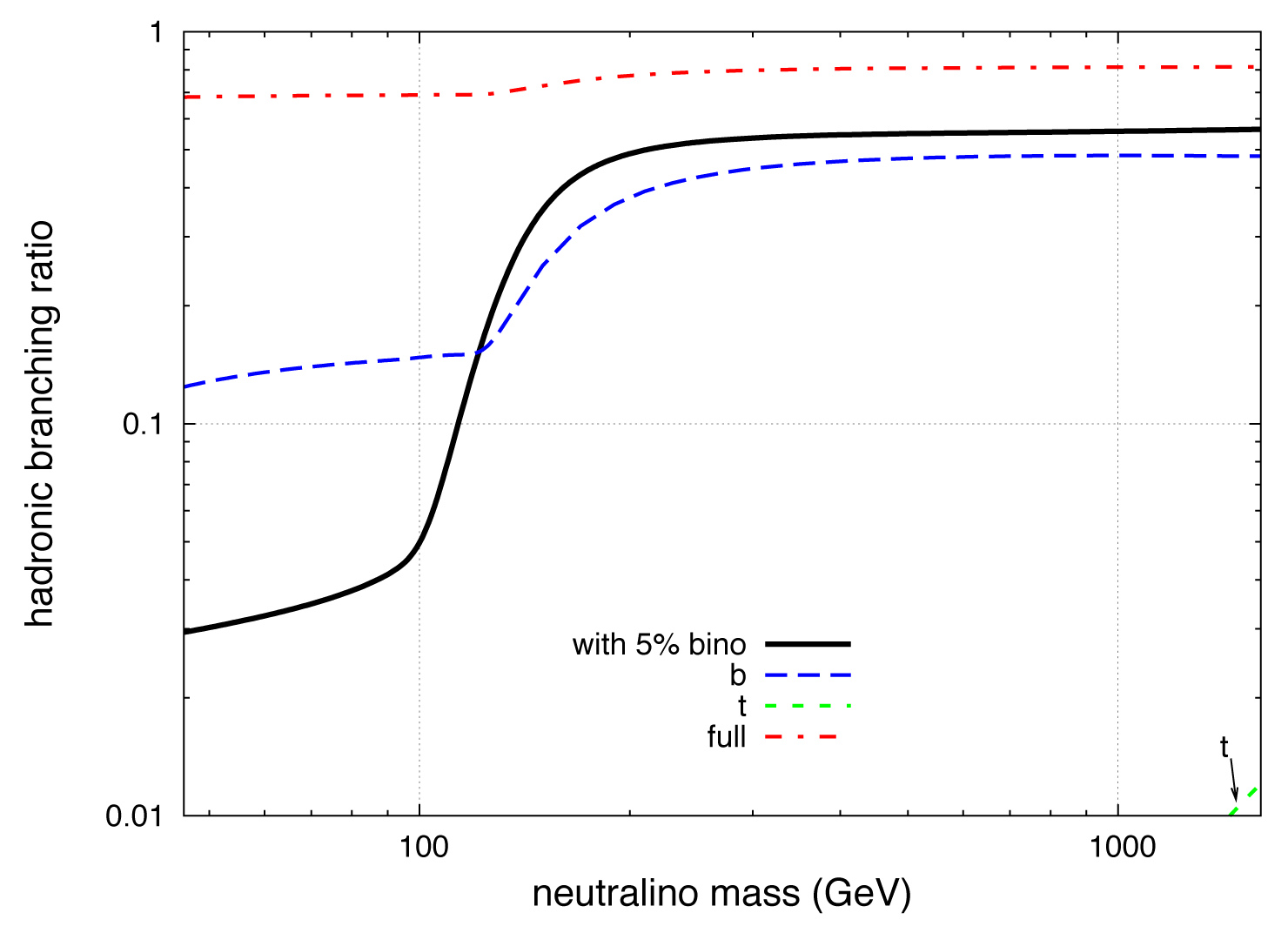}
  \caption{\small Left Panel: Branching ratio into a quark pair and a
    gravitino for a Higgsino neutralino in the limit of heavy Higgs
    decoupling and a SM-like light Higgs.  We have taken here 
    $m_h = 115$~GeV and negligible gravitino mass in the phase space factor.
    Right Panel: As the left panel, but showing the $ b\bar b $ and $ t
    \bar t $ channels and the total hadronic branching ratio with a 5\%
    Bino fraction. Below the opening of the intermediate Z/Higgs channel,
    a small Bino fraction can reduce the hadronic branching ratio very
    effectively.  }
  \label{BR-Bino-Higgsino}
\end{figure}

While the hadronic branching ratio for on-shell production is well 
approximated using the Z or $h$ hadronic widths, below these thresholds 
the expressions are more involved and there the 3-body decays 
are the dominant decay channels for a pure Higgsino. Clearly
even in that limit the branching ratio into quarks is large, as 
shown in the left Panel of Figure~\ref{BR-Bino-Higgsino}.
The expression for the decay into a b quark pair and a gravitino
is given, in the limit of negligible final state masses in the phase 
space and intermediate boson momenta in the propagators, by
 \begin{eqnarray}
 \Gamma \left( \tilde H \rightarrow \psi_{3/2} b \bar b \right) 
&=& \frac{\left| - N_{13} \cos \beta + N_{14} \sin\beta \right|^2}{48 (4 \pi)^3 M_P^2} 
\frac{m_{\tilde H}^5}{x_{\frac{3}{2}}^2} \frac{ g_Z^2}{M_Z^2} 
\left[ \left( \frac{1}{4} - \frac{2}{3} \sin^2\theta_W + \frac{8}{9} \sin^4\theta_W \right)
\right. \nonumber \\ 
 & & \left.
+ \frac{3}{10} \frac{\cos^2\alpha}{\sin^2\beta}
\frac{\left| - N_{13} \sin \alpha + N_{14} \cos\alpha \right|^2}
{\left| - N_{13} \cos \beta + N_{14} \sin\beta \right|^2} 
\frac{ m_b^2 m_{\tilde H}^2}{m_h^4}\right] \; ,
\label{Higgsino-uu}
\end{eqnarray}
where the first term comes from the intermediate Z and the second from
the light Higgs. Note that there are no favourable interference
effects, so the Higgsino always has a large hadronic branching ratio.
On the other hand the presence of even a very small gaugino fraction $
\sim 0.01-0.05 $ already makes the electromagnetic two body decay
dominant, effectively suppressing the relative size of the hadronic
decays. This effect can be seen in the right panel of
Figure~\ref{BR-Bino-Higgsino}.

Since in the Higgsino case all processes with intermediate squarks are
additionally suppressed by Yukawa couplings compared to the gaugino
case, these are surely negligible as soon as the Z or h channel opens
up. Below these thresholds, squarks that are nearly mass degenerate
with a pure Higgsino would further increase its hadronic branching
ratio, but again for mass degeneracies such that also coannihilation 
effects become important.

We summarise the results of this section in the following way: the 
neutralino usually has a large hadronic branching ratio above the 
threshold for Z production. The only way to keep the hadronic
branching ratio low for large masses is to engineer a scenario with a 
photino neutralino. Then the hadronic branching ratio
remains below 5-6\% up to 2 TeV masses. We will explore in the
following if this may help in relaxing the BBN constraints.

\medskip

\section{Neutralino thermal abundance and BBN}
\label{parameterScans}

The neutralino is considered to be one of the most promising Dark
Matter candidates since it is an explicit realisation of the WIMP
mechanism.  It was observed long ago~\cite{WIMP} that a particle with
weak interactions and mass at the electroweak scale decouples from the
thermal bath with an energy density near to the critical density.  In
general this density is approximately given by the thermally averaged
annihilation cross-section as
\begin{equation}
\Omega_{WIMP} h^2 \simeq
\frac{1\;\mbox{pb}}{\langle \sigma v \rangle} \; .
\end{equation}
While this coincidence between the weak scale and DM density is very
intriguing and suggestive, realising explicit scenarios with a
neutralino WIMP has become increasingly difficult, especially since
the accelerator bounds have now pushed the supersymmetric spectrum
higher and higher.  Nowadays we know that there is much more to the
neutralino number density than the approximate formula above and very
detailed computations and numerical programs have been developed to
take into account all annihilation channels and give reliable
predictions in the more sensitive regions such as those near
resonances or where there are coannihilations with slightly heavier
states.  In our analysis we will use one of these numerical packages,
Micromegas 2.2 \cite{micromegas}, to compute the NLSP number density
before decay.

The key quantities in the computation of the thermal abundance are the
neutralino composition in gaugino/Higgsino eigenstates and the mass
spectrum of the heavier superpartners.  To keep our analysis as
general as possible we will not fix all supersymmetric parameters
according to a specific scenario, but instead we set the soft SUSY
breaking parameters at the low energy scale. We keep the majority of
the parameters fixed and vary the gaugino and Higgsino soft mass
parameters to study how the lifetime and number density vary with the
mass and composition of the lightest neutralino. 
We use Softsusy 2.0 \cite{softsusy} to compute the physical mass spectrum 
from the soft SUSY breaking parameters. We have implemented a full 
numerical computation of the lifetime calculation \cite{JH} and have
interfaced this with the SUSY spectrum generated by Softsusy and
Micromegas. This allows us to take the results we discussed in the 
previous section and apply them to general MSSM
spectra to see how they fare against the BBN bounds.

\subsection{Bino-Wino NLSP}

\begin{figure}[t]
  \centering
  \includegraphics[scale=0.7]{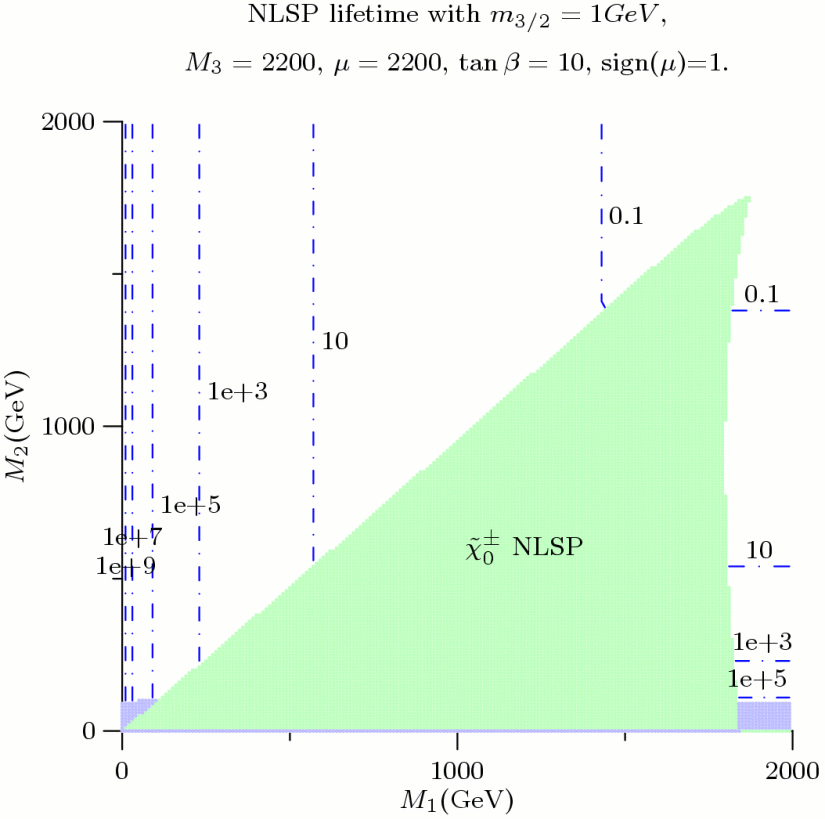}
  \caption{\small Parameter space of a Bino-Wino neutralino in the
    $M_1,M_2$ plane. $M_{1,2}$ refer here to the EW scale soft SUSY
    breaking gaugino masses. The violet (mid grey) region is excluded by
    LEP searches for the chargino, while the green (light grey) region
    has a chargino NLSP. This is because a pure Wino chargino will be
    lighter than a Wino neutralino. Bino mixing in the neutralino or
    Higgsino mixing in the chargino is required for the neutralino to be
    lighter than the chargino. A chargino NLSP is not ruled out
    physically, but we do not consider it here as we are interested
    instead in a neutralino NLSP. In the remaining parameter space we
    calculate the lifetime of the neutralino NLSP for a gravitino mass 
    of 1 GeV. The dot-dashed contours show the variation in the
    lifetime in seconds over the parameter space.} 
  \label{Bino-Wino}
\end{figure}

We first consider the scenario of a mixed Bino-Wino LSP. To study this
scenario we set $\mu$ to be much larger than $ M_1, M_2$, effectively
decoupling the Higgsinos from the gauginos. We fix the masses of the
sfermions to be above $2$ TeV and then vary $M_1$ and $M_2$ between
0 to 2 TeV. The result of this parameter scan is shown in 
Figure~\ref{Bino-Wino}.

The mixing changes continuously from practically pure Bino in the left
top corner to an equal mixture of Bino-Wino along the diagonal with
$M_1=M_2$ to nearly pure Wino in the bottom right corner.  Note though
that a small component of Higgsino is always present since we consider
a finite value for $ \mu $.
The lifetime of the neutralino in seconds for a gravitino mass of 1 GeV
is given by the dot-dashed contours in Figure~\ref{Bino-Wino}; it
is set by the physical neutralino mass and therefore the contours 
run parallel to the smaller parameter between $ M_1 $ and $ M_2 $.

\begin{figure}[ht!]
\centering
\includegraphics[scale=0.85]{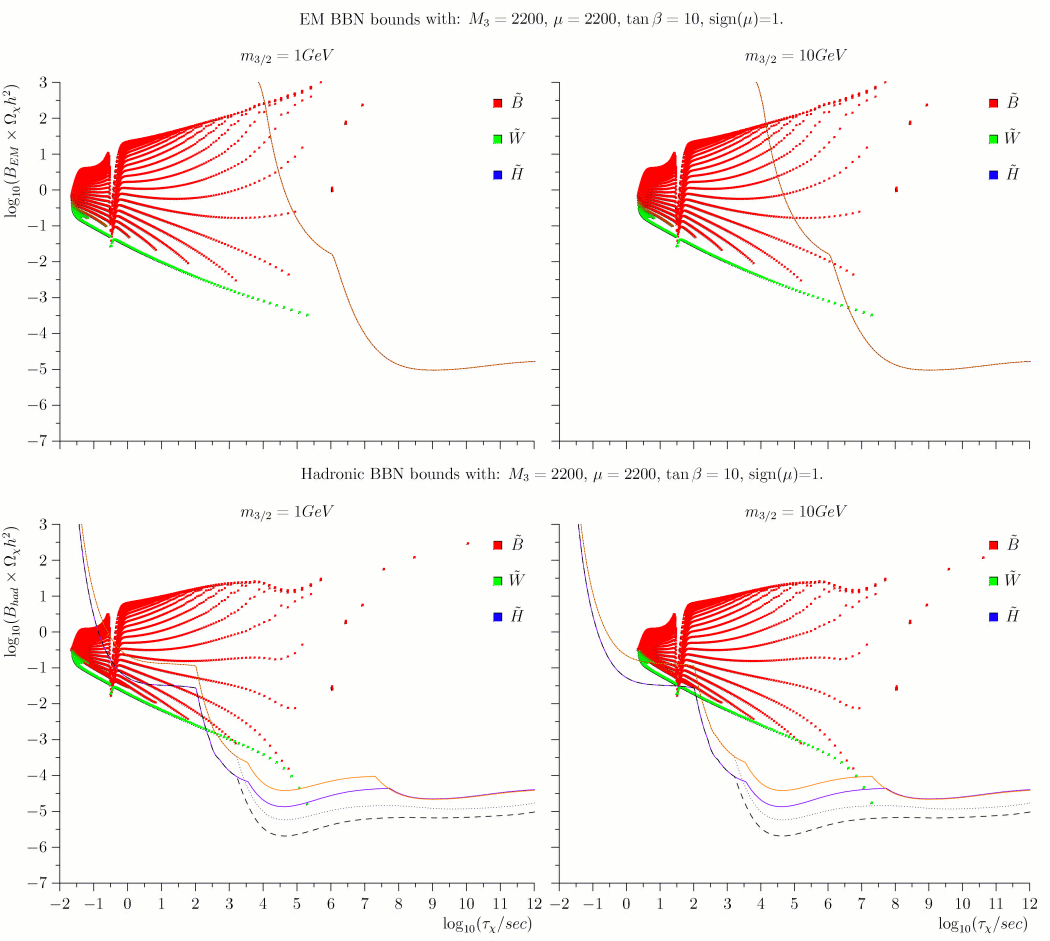}
\caption{\small Energy density of the Bino-Wino neutralino compared
  with the BBN electromagnetic (top) and hadronic constraints (bottom)
  for the case of a 1 (left) and 10 (right) GeV gravitino mass.  The
  bounds are taken from~\cite{Jedamzik}: the continuous (dashed) lines
  correspond to more (less) conservative bounds for the $^6$Li to
  $^7$Li ratio, and the region between the curves should not be
  considered as strictly excluded.  The red/upper and violet/lower
  curves in the hadronic plots are the constraints for 1~TeV and
  100~GeV decaying particle mass respectively. The points correspond
  to the allowed region from Figure~\ref{Bino-Wino}; the constant
  lifetime contours in that Figure serve as orientation to infer the
  neutralino mass and the mass parameters for the present points. The
  mass increases from right to left as heavier particles decay
  faster. The composition goes from Bino at the top to Wino at the
  bottom, and the colours give the dominant component. Note also that
  the gap between the Bino and Wino neutralino points is due to the
  presence of the region with a chargino NLSP shown in the
  Figure~\ref{Bino-Wino}.  The deformation between the upper and lower
  panels is due to the mass dependence of the hadronic branching ratio
  with lighter NLSPs having lower branching fractions to hadrons. In
  contrast the electromagnetic branching ratio is always nearly one.}
\label{BinoWino-BBN}
\end{figure}

We take all the points that satisfied our conditions i.e. all those
with a neutralino NLSP that are not ruled out by LEP bounds, and we
calculate the decay lifetime and hadronic branching fraction. 
We plot these points against the hadronic and electromagnetic BBN 
bounds in Figure~\ref{BinoWino-BBN}. The bounds are taken from the
analysis of \cite{Jedamzik} and the different curves are explained 
in the figure caption. The vertical axis corresponds to the fraction 
of the number density that decays to electromagnetic or hadronic 
products, which we approximate as the number density
of the NLSP after freeze-out multiplied by the appropriate branching
fraction.

As we can see from Figure~\ref{BinoWino-BBN} the hadronic bounds are
generally more constraining, but the electromagnetic bounds remain
important. This is especially true for light NLSP masses that, as
we have discussed, result in a low branching fraction to hadrons. 
A very light dominantly Wino NLSP almost evades the hadronic bounds 
for any gravitino mass thanks to the low density and low hadronic 
branching ratio, but does not overcome the electromagnetic bounds 
for gravitino masses of 10 GeV. The situation is better for smaller
gravitino masses and indeed for $ m_{\frac{3}{2} } = 1$ GeV 
there is a small window for a very light Wino right on the threshold 
of the LEP bounds that is still allowed even though it has a lifetime 
greater than $10^5$s.

Another important feature visible in Figure~\ref{BinoWino-BBN} is 
the large dip that corresponds to resonant annihilation into the 
pseudo-scalar Higgs, which happens for our choice of parameters for 
$ m_{\tilde \chi} \sim 1150 $ GeV.  This channel is actually open 
only thanks to a very small Higgsino component in our gaugino 
neutralino of less than 0.1~\%.
Nevertheless in the resonant region the annihilation is very
efficient, and so it dominates over all Bino annihilation
channels. Still it is not enough to avoid the constraints for large
lifetimes.  The pseudo-scalar Higgs boson has quite a large width and
cannot enhance the annihilation by more than a factor approximately
given by $\Gamma_A/M_A$.  The light Higgs resonance is more 
effective, but it is not visible in a plot with this resolution.
Also since we chose a Higgs mass of 115 GeV, the light Higgs resonance
requires a neutralino mass around 57 GeV, which is excluded for Wino
(and Higgsino) neutralinos thanks to the LEP chargino bound at around
100 GeV \cite{LEP-chargino}. Such a light Bino NLSP will always have 
a large lifetime for these values of the gravitino mass and will be
ruled out by the BBN bounds.

\begin{figure}[t]
\centering
\includegraphics[scale=0.7]{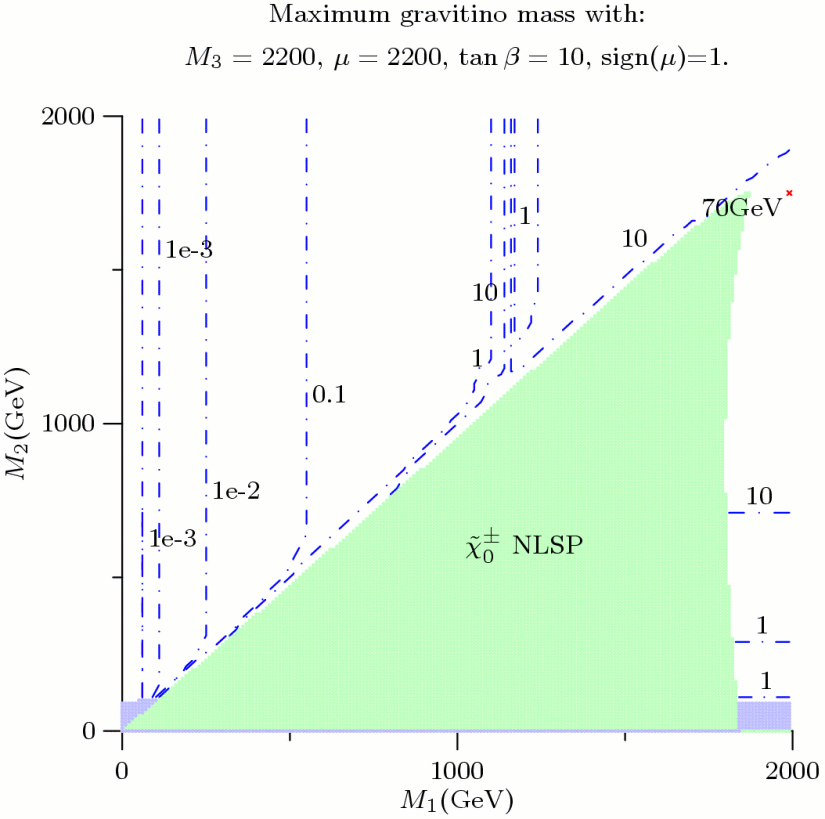}
\caption{\small Contours of maximal gravitino mass in GeV in the $ M_2
  $ vs $ M_1$ plane.  The maximal value of 70~GeV is reached at the
  boundary of our parameter space for a heavy Wino neutralino of about
  2~TeV mass, but values around and slightly above 10~GeV are possible
  also at the Higgs resonance region for a neutralino mass of 
  $ m_{\tilde \chi} \sim 1150 $~GeV, independently of the Wino-Bino 
  composition, or along the upper edge of the chargino NLSP region 
  where coannihilation helps reduce $ \Omega_{NLSP} h^2 $. 
  Also on the right side there is a window with light Wino, allowing 
  gravitino masses around 1~GeV.
  }
\label{BW-Gmass}
\end{figure}

For a gravitino mass of 1 GeV we see that a nearly pure Bino is
excluded up to the Higgs resonance at $M_1=1150$~GeV. Conversely a
dominantly Wino NLSP avoids the most stringent hadronic BBN bounds
through its small number density. This places a lower bound on the
allowed Wino mass of 300~GeV for a 1~GeV gravitino, apart for
the aforementioned small window right at the LEP mass bound. 
On the other hand for a 10 GeV gravitino mass, the only allowed 
cases are either the resonantly annihilating Bino via a small
Higgsino component or a Wino in the mass window 800-2000~GeV 
in the corner between Helium and Deuterium bounds corresponding
to a lifetime of around 100 s.  
Note that for large neutralino masses only the weaker hadronic 
constraints apply. The bounds are weaker for heavier particles since 
for constant energy density their number density is reduced by the 
inverse of their mass, while the effect of the hadronic showers 
grows more slowly than the energy released in the decay.

For larger gravitino masses the points move to the right and more 
parameter space becomes excluded. 
In Figure~\ref{BW-Gmass} we take each point in the parameter space 
and find the largest allowed gravitino mass for the given gaugino mass. 
We find the maximal gravitino mass for a gaugino neutralino below 2~TeV 
in mass to be 70~GeV and this occurs for a heavy Wino with a small 
but non-zero Bino fraction. We also see that a Wino NLSP can allow 
a gravitino of a few GeV for an NLSP mass of a few hundred GeV, 
or tens of GeV for an NLSP with a mass of over 750~GeV. 
On the right of the chargino NLSP region, we find the very light
Wino, on the very edge of the LEP bound, which still allows for
a gravitino with a mass of a few GeV.

The Bino case is substantially more restrictive. The only regions 
that allow for a gravitino with a mass over a GeV are in the regions 
with a mixed Bino-Wino (along the diagonal in the $(M_1,~M_2)$ plane), 
in the pseudo-scalar Higgs resonance which is here around 
$M_1=1150$~GeV, or for very large Bino masses. 
In the majority of cases, the rise in the maximum gravitino
mass is due to a reduction in the NLSP number density after freeze-out
through enhanced annihilation either through the Higgs resonance or
along the chargino NLSP region where coannihilation is effective. 
The reduction of the hadronic branching fraction, e.g. for the 
fine-tuned photino case, is not sufficient to relax the constraints;
the only exception of this is the light Wino case on the LEP boundary,
which has both a reduced number density thanks to the strong Wino
annihilation and a small hadronic branching ratio around the
threshold for intermediate on-shell Z.

\subsection{Bino-Higgsino NLSP}

\begin{figure}[ht!]
\centering
\includegraphics[scale=0.7]{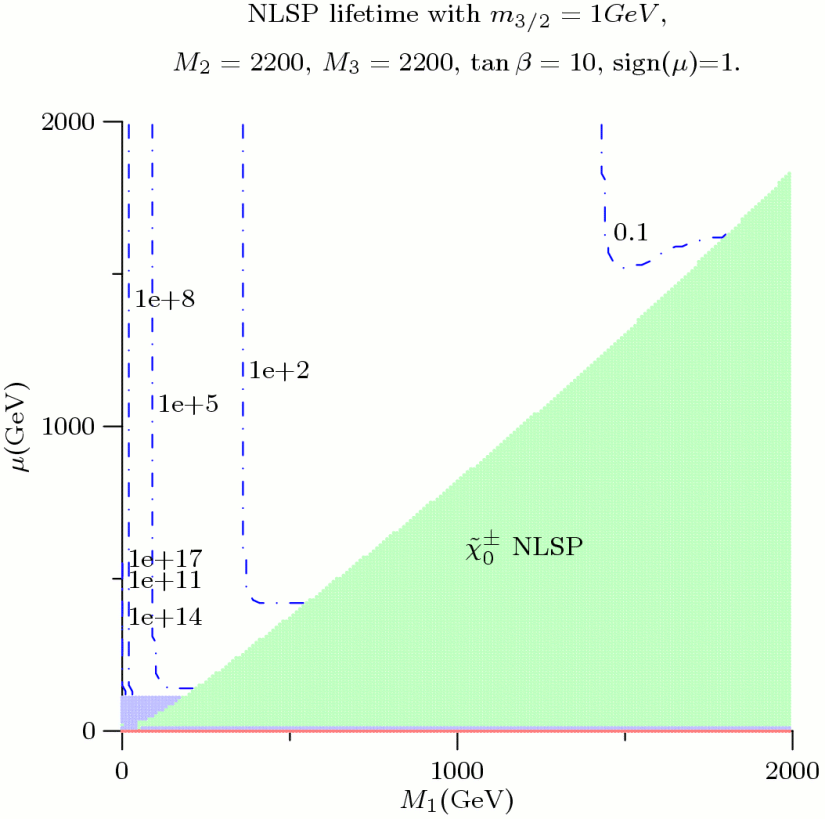}
\caption{\small Parameter space of a Bino-Higgsino neutralino.  The
  violet (mid grey) region is excluded by LEP searches, while the
  green (light grey) region has chargino NLSP.  The contours give the
  neutralino lifetime in seconds for a gravitino mass of 1~GeV. 
  The lifetime for other values of the mass is just rescaled as
  $\left(m_{\frac{3}{2}}/1~\mbox{GeV} \right)^2 $ as long as phase space
  factors are negligible.}
\label{Bino-Higgsino}
\end{figure}

\begin{figure}[ht!]
\centering
\includegraphics[scale=0.85]{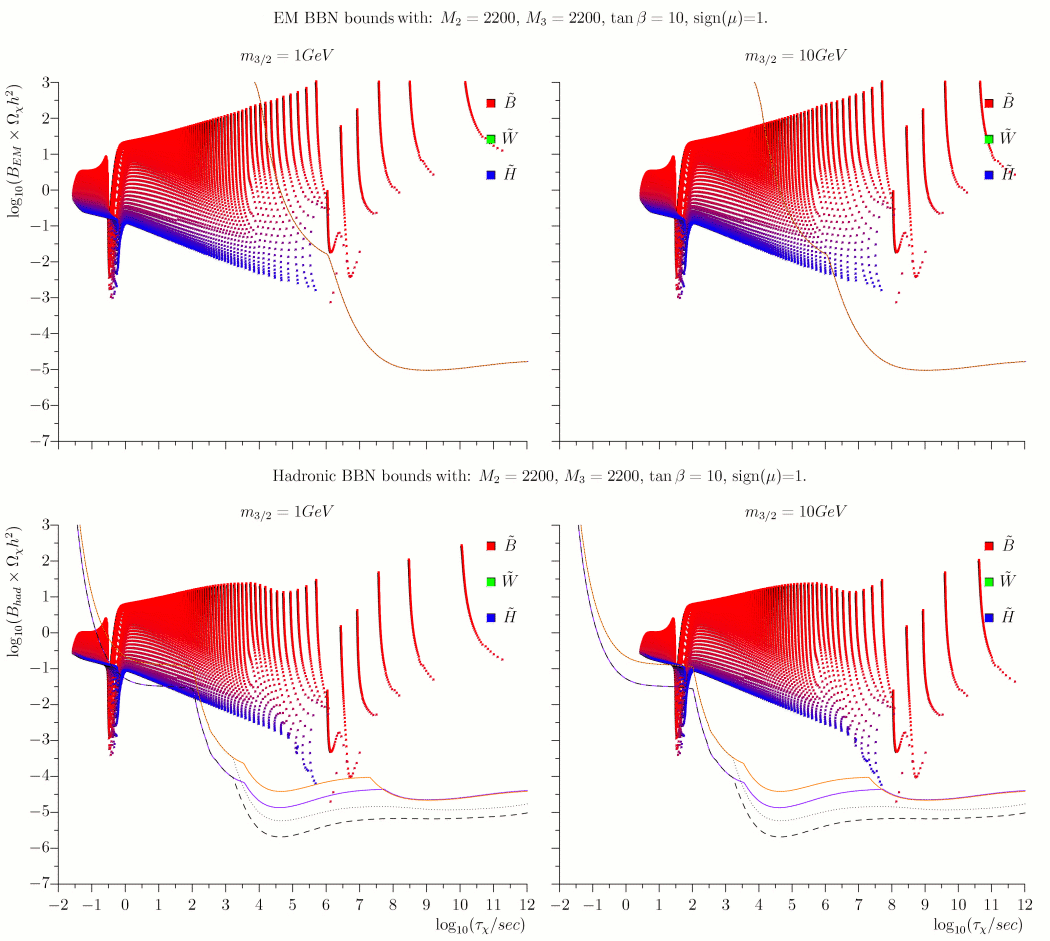}
\caption{\small Energy density of the Bino-Higgsino neutralino
  compared with the BBN electromagnetic (top) and hadronic (bottom)
  constraints for the case of a 1 (left) and 10 (right) GeV gravitino
  mass.  The bounds are taken from \cite{Jedamzik}: the continuous
  (dashed) lines correspond to more (less) conservative bounds for the
  $^6$Li to $^7$Li ratio. The region between these curves should not
  be considered as strictly excluded, but there a substantial Lithium
  abundance arises from the NLSP decay.  The red/upper and
  violet/lower curves in the hadronic plots are the constraints for
  1~TeV and 100~GeV decaying particle mass respectively.  The points
  correspond to allowed points from Figure~\ref{Bino-Higgsino}; the
  constant lifetime contours in that Figure serve as orientation to
  infer the neutralino mass and the mass parameters for the shown
  points. The neutralino mass increases from right to left and its
  dominant component is given by the colour coding as indicated.  The
  deformation between the upper and lower panels is due to the mass
  dependence of the hadronic branching ratio, while the
  electromagnetic branching ratio is always nearly one.}
\label{BinoHiggsino-BBN}
\end{figure}

In order to study the Bino-Higgsino case we consider a scenario with
large $M_2$ parameter and study the $ \mu, M_1 $ plane as shown in
Figure~\ref{Bino-Higgsino}.  The other supersymmetric scalar particles
are heavy at around 2.2 TeV.
Note that in this case nearly all the parameter space with $ \mu < M_1 $ 
corresponds to a Higgsino chargino NLSP and will not be considered here.
Therefore we never have a pure Higgsino NLSP and the region with
mostly Higgsino composition along the diagonal is characterised 
by a nearly mass degenerate Higgsino chargino.

\begin{figure}[t]
\centering
\includegraphics[scale=0.7]{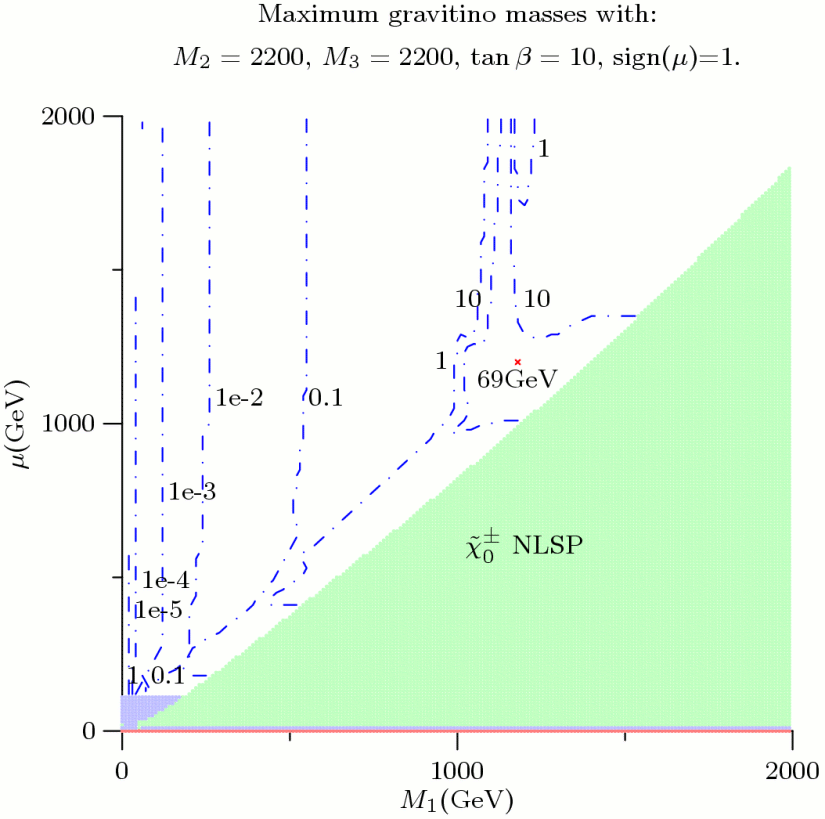}
\caption{\small Contours of maximal gravitino mass in GeV in the $ \mu
  $ vs $ M_1$ plane. The maximal value of 69~GeV is reached at the
  heavy Higgs resonance. There is also a thin band at light masses that
  allows a gravitino of order a few GeV due to the effect of the light
  Higgs resonance just above the LEP bound.}
\label{BH-Gmass}
\end{figure}

The Nucleosynthesis constraints for a mixed Bino-Higgsino case are
shown in Figure~\ref{BinoHiggsino-BBN}. Each point corresponds to a
point from Figure~\ref{Bino-Higgsino} that satisfies the LEP bounds
and give a neutralino NLSP. We see here again that a light mostly 
Higgsino neutralino, as the Wino, has a suppressed hadronic branching 
ratio and can nearly evade the hadronic constraints, but not the 
electromagnetic bounds.  
In this case the resonant annihilation, both via the pseudo-scalar and 
via the heavy scalar Higgs, nearly mass degenerate for our choice of 
parameters, proceeds much more efficiently thanks to the large Higgsino 
fraction and is thus more prominent.  This resonant annihilation 
overcomes the BBN constraints up to gravitino masses of the order of 
70~GeV, as can be seen clearly in Figure~\ref{BH-Gmass}. 
In this case the coannihilation with the lightest Higgsino chargino 
serves to keep the Higgsino number density small along the chargino
NLSP boundary and consequently results in a larger maximum gravitino
mass. 
However the largest reduction of the number density comes from
the heavy Higgs resonance and the rest of the parameter space is
excluded for neutralino masses below 2~TeV and gravitino masses above
10~GeV.

\newpage

\subsection{Wino-Higgsino NLSP}

\begin{figure}[t]
\centering
\includegraphics[scale=0.7]{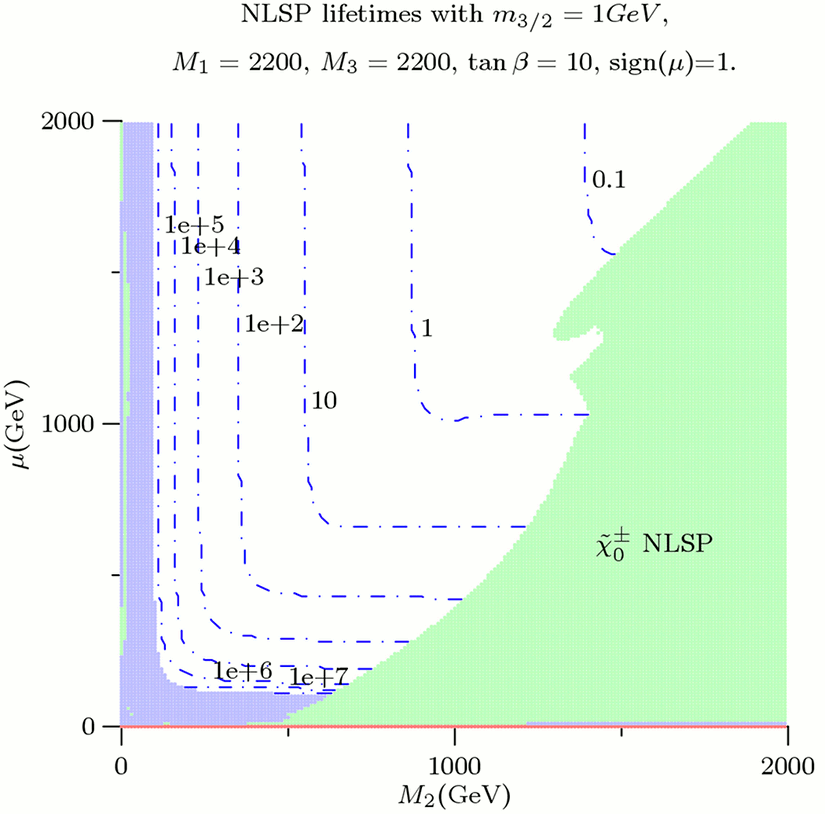}
\caption{\small  Parameter space of a Wino-Higgsino neutralino.  
The violet (mid grey) region is excluded by LEP searches, while the 
green (light grey) region has a chargino NLSP. 
The contours give the neutralino lifetime in seconds for a gravitino 
mass of 1~GeV. The lifetime for other values of $ m_{\frac{3}{2}} $ 
is just rescaled as $\left(m_{\frac{3}{2}}/1~\mbox{GeV} \right)^2 $ 
as long as phase space factors are negligible.}
\label{Wino-Higgsino}
\end{figure}

The Wino-Higgsino case is best considered in the case of a large $M_1$
parameter and then the parameter space is given in the  $ \mu,
M_2 $ plane, as shown in Figure~\ref{Wino-Higgsino}.  The other
supersymmetric particles are as heavy as in the previous cases.
Also in this case a pure Higgsino is not present in the allowed
parameter space, but the Higgsino fraction can be larger than
in the Bino-Higgsino case.

\begin{figure}[ht!]
\centering
\includegraphics[scale=0.9]{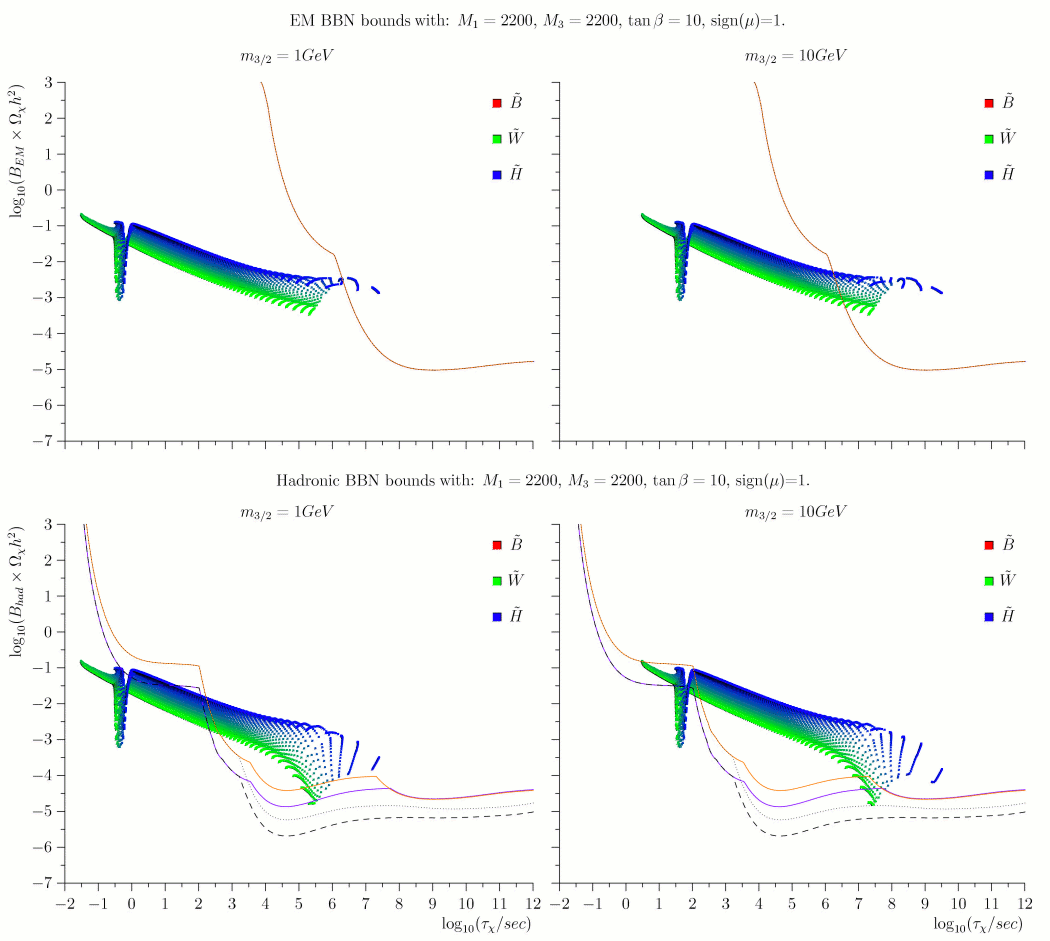}
\caption{\small Energy density for a Wino-Higgsino neutralino compared
  with the BBN electromagnetic (top) and hadronic (bottom) constraints
  for the case of a 1 (left) and 10 (right) GeV gravitino mass.  The
  bounds are taken from \cite{Jedamzik}: the continuos (dashed) lines
  correspond to more (less) conservative bounds for the $^6$Li to
  $^7$Li ratio, and the region between these curves should not be 
  considered as strictly excluded, but there a substantial
  Lithium abundance arises from the NLSP decay.  The red/upper and
  violet/lower curves in the hadronic plots are the constraints for 
  1~TeV and 100~GeV decaying particle mass respectively. 
  The points correspond to allowed points from Figure~\ref{Wino-Higgsino};
  again the constant lifetime contours in that Figure serve as orientation
  to infer the neutralino mass and the mass parameters 
  for the present points. The NLSP mass increases from right to left 
  and the dominant component is given by the colour coding. 
  The deformation between the upper and lower panels is due to the 
  mass dependence of the hadronic branching ratio, while the 
  electromagnetic branching fraction is always nearly one.  }
\label{WinoHiggsino-BBN}
\end{figure}

The Nucleosynthesis constraints for a mixed Wino-Higgsino case are
shown in Figure~\ref{WinoHiggsino-BBN}.  As before, each point
corresponds to a point from Figure~\ref{Wino-Higgsino} that satisfies
LEP bounds and gives a neutralino NLSP. The number density in this
case is substantially smaller than for a Bino neutralino, and mostly
lower than the DM density for a neutralino mass below 2~TeV, but not
sufficiently to avoid all the bounds. We see here again that a light
Wino-Higgsino has suppressed hadronic branching ratio and can evade
all the hadronic constraints, especially if the gravitino mass is
sufficiently large to give a suppression of the Z channel.  This
effect is most pronounced for 10~GeV gravitino masses since the
lightest Wino-Higgsino mass is around 100~GeV. However in the case of
a 10~GeV gravitino, the light neutralino regions of parameter space 
are excluded again by the electromagnetic constraints. As in the
Wino-Bino case, we see here that there is a small window for a Wino 
with a mass just above the LEP bound and a gravitino of around 1~GeV. 
However in most cases in the low mass region, below the Z resonance, 
the interplay of the two types of bounds exclude a light neutralino NLSP.

\begin{figure}[t]
\centering
\includegraphics[scale=0.7]{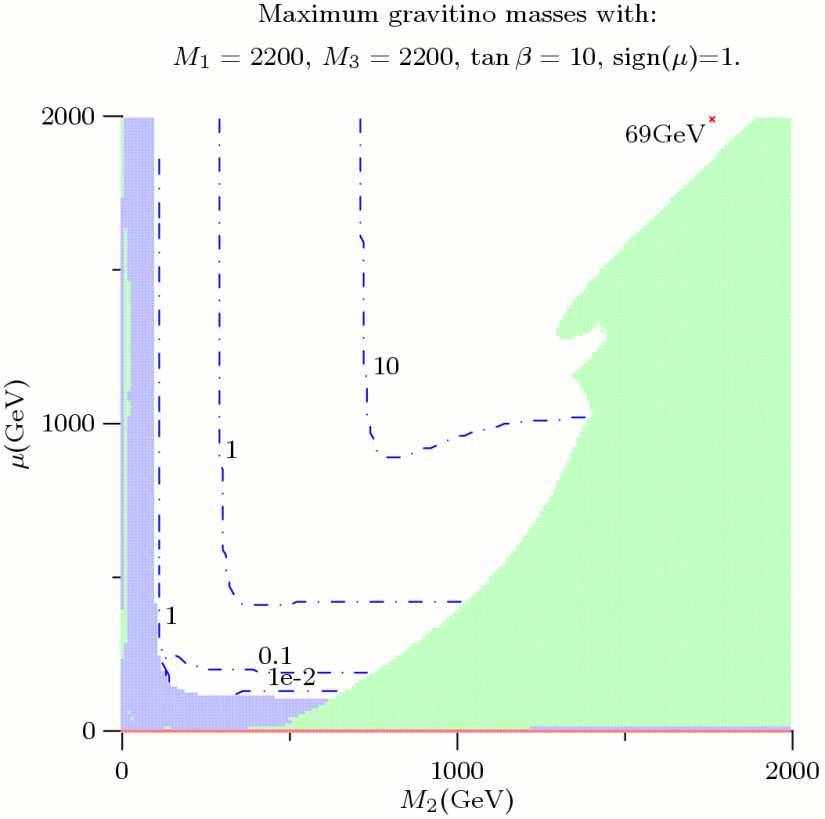}
\caption{\small Contours of maximal gravitino mass in GeV in the $ \mu
  $ vs $ M_2 $ plane. The maximal value of 69~GeV is reached again at
  the boundary of the parameter space for a neutralino mass of about 
  2~TeV. We also see a thin band at $M_2\approx 100$~GeV where the low
  hadronic branching fraction allows a gravitino mass up to a few GeV.
}
\label{WH-Gmass}
\end{figure}

In this case, for gravitino masses of 10 GeV, apart from the resonant
annihilation case at 1150 GeV neutralino mass, once again there is a
surviving window for an NLSP mass of 800-2000 GeV in the corner between
Helium and Deuterium bounds.  This window remains open up to gravitino
masses of about 70~GeV.  It is not surprising that we find a very
similar maximal value for all our cases, since in two of them it is
located at the boundary of our parameter region, where the neutralino
is actually a fully mixed state.  Note however that in the case of a
Wino-Higgsino neutralino there is a wide region of allowed gravitino
masses between 10 and 69~GeV, for neutralino masses above 800 GeV,
which is not limited to the Higgs resonance. 
This is shown in Figure~\ref{WH-Gmass}.
So we can conclude that the BBN constraints are strongly relaxed
for the Wino-Higgsino case thanks to the low relic density.

\section{General neutralino NLSP}

Combining the results of the previous sections we can obtain the
general result for any neutralino composition.
We observe the following features:

\begin{itemize}

\item{The reduction of the hadronic branching ratio is more
effective for low neutralino masses, thanks to the phase space
suppression for the Z and Higgs channels, and can help to
overcome the BBN bounds for Wino or Higgsino neutralinos.
Unfortunately in most cases the electromagnetic constraints
are not bypassed.  This effect does not allow for gravitino
masses above few GeV.
}

\item{
The suppression of the hadronic channels due to a photino composition 
is not sufficient to relax the hadronic constraints,  since 
the photino relic density is always too large. Moreover for low masses 
and large lifetimes the photino is excluded by the 
electromagnetic bounds.}

\item{
The resonant Higgs annihilation can lower the relic density sufficiently 
to allow for gravitino masses up to 70 GeV for our choice of heavy Higgs 
masses. This maximal value comes from the point at which the resonant 
spike touches the BBN bounds at around $10^{2.5-3} $ s and therefore 
for the general case the maximal gravitino mass is given approximately 
by the neutralino mass as
\begin{equation}
\left(m_{3/2} \right)_{max, res} \sim
70\; \mbox{GeV} 
\left( \frac{m_{\tilde \chi}}{1.15 \mbox{TeV}} \right)^{5/2}\; .
\end{equation}
The actual value is therefore smaller for lower neutralino and heavy 
Higgs masses and cannot allow for 10 GeV gravitino mass for neutralino 
masses below 500 GeV.
}

\item{
The coannihilation with the charginos and the stronger
    annihilations characteristic of Wino and Higgsino NLSPs result in
    a lower number density of NLSPs in the early universe. This helps
    evade the strongest BBN bounds and results in larger allowed
    gravitino masses than for a predominantly Bino NLSP.}

\item{ 
In the previous sections we kept the sfermions heavy and focused on 
the effect of the neutralino composition. In many cases we would expect 
the sfermion masses to be close in mass to the lightest neutralino, 
particularly in cases with a gaugino NLSP and light sleptons. 
Then sfermion coannihilation can significantly reduce the number density 
of a neutralino NLSP. 
We studied the sfermion coannihilation case and found it to have a
significant effect on the Bino NLSP number density, allowing an order 
of magnitude increase in the maximum gravitino mass. 
Conversely, in the case of a dominantly Higgsino or Wino NLSP, the number 
density is only slightly altered by sfermion coannihilation and the 
resulting change in the number density has little impact on the allowed 
gravitino mass.
We show the effect of coannihilation with a light stau for
a Bino NLSP in Figure~\ref{coann} and we obtain qualitatively
similar results for a stop NNLSP.
}

\begin{figure}[ht!]
\centering
\includegraphics[scale=0.83]{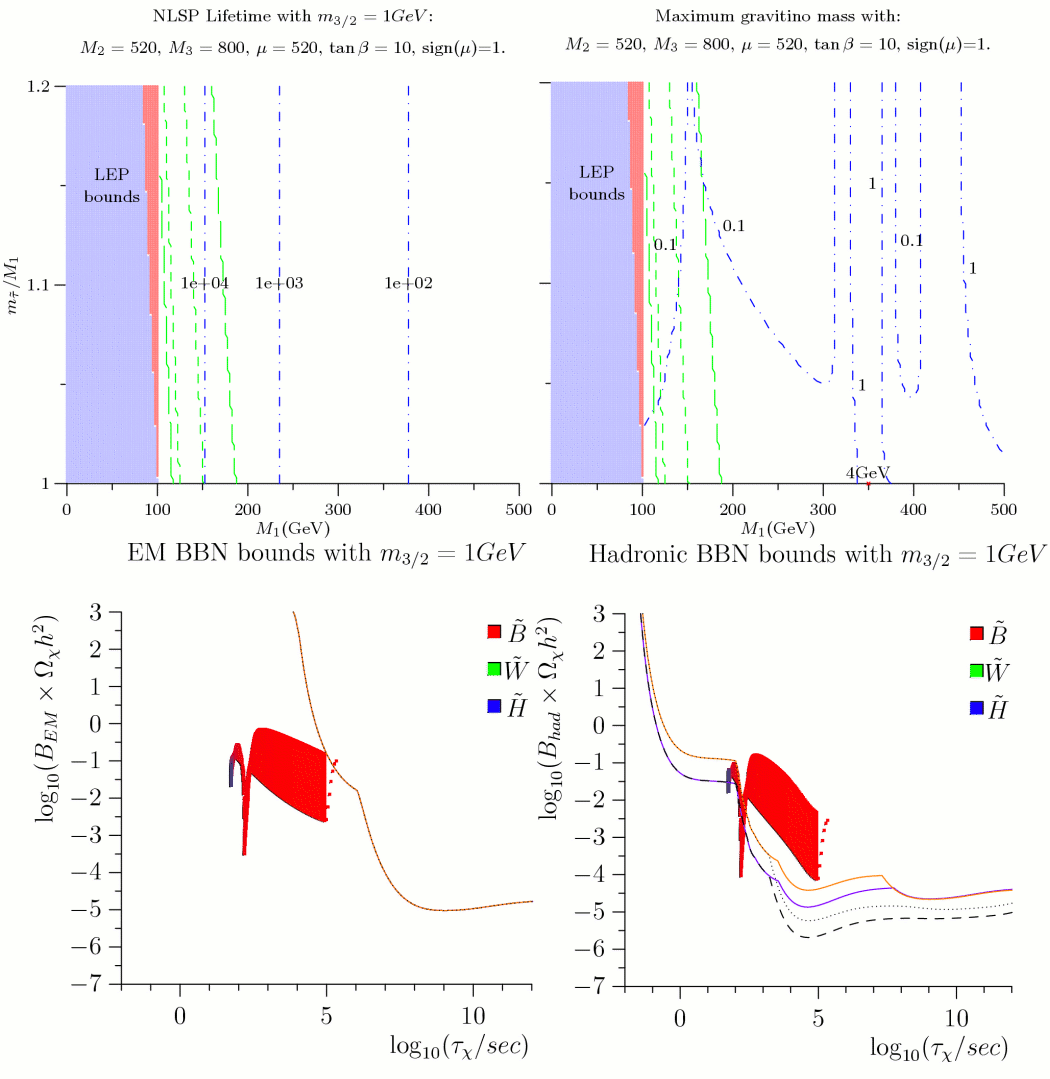}
\caption{\small 
The effect of coannihilation with a light stau for a Bino NLSP. 
In the top two plots we show the parameter space: on the horizontal axis 
we take the increasing Bino mass, while the vertical axis shows the mass
splitting between the neutralino NLSP and the stau NNLSP. 
The top left plot shows the lifetime contours for a 1~GeV gravitino
in seconds. Note that the lifetime of the neutralino is dependent on 
the neutralino mass, but insensitive to the stau mass. 
The top right plot shows the maximum gravitino mass across the parameter 
space. The overall behaviour is that the maximum mass increases with 
increasing neutralino mass and shortening lifetime. 
The exception to this behaviour is around 350~GeV where we hit the Higgs 
resonance, here we have taken the pseudo-scalar Higgs mass to be 700~GeV. 
We also see an increase in the maximum mass as the mass splitting 
approaches zero. Here the number density is reduced enough to avoid 
the $^4$He bound, corresponding to lifetimes below 100 s.
This results in an order of magnitude increase in the maximum allowed
 gravitino mass.}
 \label{coann}
\end{figure}

\item{
We have also studied the effect of changing the value of $\tan\beta$. 
This has little impact on the lifetime calculations, instead having the 
most direct effect through the location of the Higgs resonances for 
the NLSP annihilations.
}

\end{itemize}

We see therefore that a general neutralino NLSP extends the
range of gravitino masses allowed by about one order of magnitude
compared to the Bino case. Our study shows that the smallest splitting
between the allowed gravitino mass and NLSP mass occurs in the regions
of resonant annihilation or Wino-Higgsino neutralinos. In these cases
a substantial hierarchy remains necessary between the NLSP and the
gravitino mass.

\newpage

\section{Conclusions}

We have considered in this paper the BBN constraints on a general
neutralino NLSP for varying gravitino masses.  We found that the
bounds are weaker for the case of a Wino or Higgsino neutralino, but
still not sufficiently relaxed to allow a 100 GeV mass gravitino,
unless we allow for neutralino masses in excess of 2 TeV.  The maximal
gravitino mass we find for neutralinos below 2 TeV is around 70 GeV,
mostly at the boundary of the parameter space.  Nevertheless gravitino
masses in the 10-70 GeV window are possible, even for lower neutralino
masses, especially in regions with resonant annihilation of the
NLSP into heavy scalar or pseudo-scalar Higgs or for a Wino-Higgsino
neutralino.  These values of the gravitino mass may still be
marginally in agreement with the thermal leptogenesis scenario
\cite{leptogenesis} if the gluinos are light \cite{Fujii:2003nr} .  
For our choice of
parameters, i.e. $ M_3 = 2.2 $ TeV, they are unfortunately too heavy,
but lowering their mass should not modify our results too much,
especially in the Higgs resonance region.  Then assuming them to be
just a bit heavier than the neutralino, and a gravitino mass of 70
GeV, the reheat temperature for gravitino Dark Matter is given by
\begin{equation}
T_R \sim 1.5  \times 10^{9} \mbox{GeV}
\left(\frac{m_{\tilde g}}{1.25\; \mbox{TeV} }\right)^{-2} \; .
\end{equation}
Since the bound $ T_R > 1.5 \times 10^9 $ GeV has been obtained for
thermal leptogenesis requiring independence from the initial
conditions~\cite{Blanchet:2006be}, our scenario may still be
acceptable, within the order one uncertainty in the thermal
computation, especially with a small enhancement of the CP violation.
This low value of the gluino mass would again call for non universal
gaugino masses at the GUT scale.

Another open window in the parameter space is the Wino NLSP region
just above the LEP bound, where kinematic effects conspire to reduce
the neutralino hadronic branching ratio. In that case the gravitino
mass is of the order of a few GeVs and the gluino mass may be
substantially smaller than 1 TeV, such as to still allow for
successful leptogenesis.

A general lowering of the $M_3$ mass parameter in comparison to the
other masses is also welcome in order to observe this scenario at the
LHC. Assuming the gluinos to be below 2 TeV of mass, the main
observable will be missing energy in the cascade decays, as for the
case of neutralino Dark Matter.  Unfortunately, the resonant
annihilation region is difficult to investigate at the LHC, since very
precise measurements of the neutralino and heavy Higgs masses are
necessary to disentangle the neutralino LSP and DM case from the one
we discuss here. In contrast, a Wino-Higgsino NLSP scenario is more
easily identified due to the existence of nearly degenerate charginos.
Nevertheless it will probably be difficult to prove that the
neutralino number density is much lower that required for dark matter
from LHC measurements alone.

Regarding the exploitation of the NLSP decay to produce the whole DM
density, it is not possible for a Wino-Higgsino neutralino in our
parameter space due to the small relic density, below 0.1.  For a Bino
NLSP, the ratio between gravitino and neutralino mass is always
smaller than $ 10^{-3} $ away from the Higgs resonance, where the
relic density is in any case suppressed. From the
Figures~\ref{BinoWino-BBN} and \ref{BinoHiggsino-BBN} we see that even
the Bino relic density for a 2 TeV mass is not sufficient to compensate a 
factor of $ 10^{-3} $ reduction in the energy density and produce the whole 
gravitino Dark Matter population.  Much heavier
NLSP neutralinos are needed to obtain the right abundance and avoid
BBN bounds for gravitino masses of 1-10 GeV~\cite{sweetspot}.  Such
large masses are unfortunately beyond the reach of the LHC and cannot
be reconciled with thermal leptogenesis.

From our analysis we can in general conclude that non-universal gaugino 
masses with a compressed gaugino spectrum, and moreover NLSP masses above 
500 GeV or so are preferred. Any evidence of a light neutralino
at LHC, apart for the case of a light Wino with nearly degenerate 
charginos, would be difficult to reconcile with gravitino Dark Matter and
leptogenesis in the standard cosmology picture with R-parity conservation.
On the other hand, heavier neutralino masses and strong enhancement of 
the NLSP annihilation as in the Higgs resonance case may be the first 
phenomenological signal for gravitino DM at colliders.

{\bf Note added:} During the completion of this work, ref.~\cite{Cyburt:2009pg}
appeared, which also computes Nucleosynthesis constraints for a general 
neutral decaying particle and in particular for a decaying gravitino. 
In \cite{Cyburt:2009pg} no BBN bound for lifetimes below 100 s
are given, due to a weaker upper limit on $ ^4$He compared
to \cite{Jedamzik}. Such weakening of the bounds would not have 
any effect on our results for the Wino-Higgsino case, since there the
relic density is below the $ ^4$He  curve, but it may open 
up more parameter space for the Bino neutralino, beyond the Higgs 
resonance region.

\section*{Acknowledgements}

LC would like to thank W. Buchm\"uller, S. Davidson, A. Romanino, M. Serone,  
F. Strumia and F. Zwirner for useful discussions on the WW decay channel.
LC would also like to thank the Aspen Center for Physics for hospitality
during the final stages of this work.  

LC and JR acknowledge financial support via the "Marie Curie Host 
Fellowship for Transfer of Knowledge" MTKD-CT-2005-029466 during their 
stay in the Institute for Theoretical Physics, University of Warsaw, 
where this work has been started.
LC acknowledges the support of the "Impuls- und Vernetzungsfond" of 
the Helmholtz Association under the contract number VH-NG-006. 
The work of JR is supported by the NSF Career Grant PHY-0449818 
and DOE OJI grant \#DE-FG02-06ER41417.

\pagebreak

\end{document}